\documentclass[galaxies,article,accept,moreauthors]{Definitions/mdpi} 
\firstpage{1} 
\pubvolume{1}
\issuenum{1}
\articlenumber{0}
\pubyear{2026}
\copyrightyear{2026}
\datereceived{ } 
\daterevised{ } 
\dateaccepted{ } 
\datepublished{ } 
\longauthorlist{yes} 

\Title{Spectroscopic orbits for three SB2s and one hierarchical triple using SALT data.}

\Author{Mikhail Yu. Kovalev $^{1*,2}$\orcidA{}, Alexey Yu. Kniazev$^{3,4,5}$\orcidB{} and Oleg Yu. Malkov $^{6}$\orcidC{}}

\AuthorNames{Mikhail Kovalev, Alexey Kniazev and Oleg Malkov}

\address{%
$^{1}$ \quad Yunnan Observatories, Chinese Academy of Sciences, Kunming, China; mikhail.kovalev@ynao.ac.cn\\
$^{2}$ \quad International Centre of Supernovae (ICESUN), Yunnan Key Laboratory of Supernova Research, Yunnan Observatories, Chinese Academy of Sciences (CAS), Kunming 650216, China\\
$^{3}$ \quad South African Astronomical Observatory, 7935 Observatory, Cape Town, South Africa;\\ 
$^{4}$ \quad Southern African Large Telescope Foundation, 7935 Observatory, Cape Town, South Africa;
a.kniazev@saao.nrf.ac.za\\
$^{5}$ \quad Sternberg Astronomical Institute, Lomonosov Moscow State University, Moscow, Russia\\
$^{6}$ \quad Institute of Astronomy, Russian Academy of Sciences, Moscow, Russia; malkov@inasan.ru}
\corres{Correspondence: mikhail.kovalev@ynao.ac.cn}

\abstract{
We confirmed four spectroscopic binary candidates using new observations obtained with SALT. Three SB2 systems (HD 20784, HD 43519A, HD 62153A) exhibit circular orbits with periods shorter than 10 days, whereas one hierarchical triple system (HD 56024) contains a close binary with an inner eccentric orbit with a period of approximately 14 days, composed of nearly identical stellar components, and a rapidly rotating star on an outer eccentric orbit with a period of approximately 400 days. For two additional SB2 candidates (HD 198174 and HD 208433), our new observations do not allow us to derive reliable orbital solutions.}


\begin{document}

\def\kms{\,{\rm km}\,{\rm s}^{-1}}
\def\Msun{\,{\rm M}_\odot}
\def\feh{\hbox{[Fe/H]}}
\def\mgfe{\hbox{[Mg/Fe]}}
\def\tife{\hbox{[Ti/Fe]}}
\def\mnfe{\hbox{[Mn/Fe]}}
\newcommand{\teff}{T_{\rm eff}}
\newcommand{\rv}{{\rm RV}}
\def\Vmic{V_{\rm mic}}
\def\Vmac{V_{\rm mac}}
\def\vsini{(V \sin{i})}
\def\logg{\log{\rm (g)}}
\def\snr{\hbox{S/N}}
\newcommand{\ha}{\hbox{H$\alpha$}}
\newcommand{\mnras}{MNRAS}
\newcommand{ \aaps}{AAPS}
\newcommand{ \aj}{AJ}
\newcommand{ \apj}{ApJ}
\newcommand{\aap}{A\&A}
\newcommand{\apjs}{ApJS}



\section{Introduction}

{ Binary and higher-order multiple stellar systems are important astrophysical objects, as approximately half of all stars in the Milky Way are not single \citep{moe2017mind}. The investigation of their orbital properties (e.g. period, eccentricity) and fundamental stellar parameters (e.g. masses, radii, effective temperatures) is essential for the development and calibration of binary population synthesis models, which in turn are required for a robust understanding of star formation processes and the subsequent evolution of binary systems \citep{han2020binary}.} If orbital orientation allow us to see periodic changes in the { radial velocities derived from spectroscopic observations of the stellar system,} they are called spectroscopic binaries (SBs). SBs are divided between single-lined (SB1), double-lined (SB2), triple-lined (SB3), and $n$-lined SB$n$ depending on number $n$ of the visible spectroscopic components. { SB2 systems are of particular significance because they permit model-independent determinations of stellar masses, provided that the orbital inclination is constrained through complementary photometric or astrometric observations \citep{gaia_dr3multiple}. SB3 systems predominantly form hierarchical configurations, in which a single star orbits an inner close binary \citep{tokovinin25,sb3japan}. Their dynamical stability, as well as the formation mechanisms, constitute key subjects of current investigation \citep{sb3sim,tokovinin2}. } 

Recently, \cite{sarah2022} observed 166 B-type stars with the FEROS spectrograph \citep{feros}, selected by variability in TESS \citep{tess} photometry, and identified 26 SB2 candidates. For one of them, HD 20784, circular orbit was found in \citep{hd20784}; however, it was based only on three spectra with poor phase coverage, so it needs to be verified by additional observations. In this paper we report results for analysis of follow-up observations for 6 of these SB2 candidates, see Table~\ref{tab:inp} with brief information. { None of the systems exhibit eclipses in their light curves, with only HD 20784 displaying ellipsoidal variability \citep{hd20784}. We confirmed that HD 20784 follows a circular orbit, derived detailed orbital solutions for two additional SB2s, HD 49519A and HD 62153A (also known as CD-73~375A), and identified a new hierarchical  system, HD 56024. For the remaining two systems, HD 198174 and HD 208433, our new observations are insufficient to derive robust orbital solutions; therefore, they must still be classified as SB2 candidates.} 


\section{Materials and Methods}

\subsection{observations}
We conducted spectroscopic observations with the Southern African Large Telescope \citep[SALT;][]{2006SPIE.6267E..0ZB,2006MNRAS.372..151O} High Resolution échelle Spectrograph \citep[HRS;][]{2008SPIE.7014E..0KB,2010SPIE.7735E..4FB,2012SPIE.8446E..0AB,2014SPIE.9147E..6TC}. Several spectra (from 5 to 13) were taken for each target using the low resolution mode ($R=\lambda/\Delta\lambda\sim$14 000) in  2023, 2024. The SALT HRS has two arms: blue 3700-5500 Å and red 5500-8900 Å. Since our targets are hot stars, with most of the spectral lines having $\lambda<5500$ Å, we used only the blue arm. The primary reduction of HRS data was produced automatically using the standard pipeline processing described in \citet{2010SPIE.7737E..25C}. The reduction of spectroscopic data for HRS was performed using the standard data processing system, detailed in \citet{2016MNRAS.459.3068K,2019AstBu..74..208K}.
Since HRS is a thermally stabilized instrument, all standard calibrations are performed weekly, { which is sufficient to achieve a radial-velocity precision of} $0.2-0.5~\kms$ depending on the observational mode.

\begin{table*}
    \centering
    \begin{tabular}{lccccccc}
    \hline
    Star& $\alpha^\circ$(ICRS)& $\delta^\circ$(ICRS)& $G$, mag& $\pi$, mas & RUWE &ST &N\\
    \hline
  HD 20784 & 49.60306 & -55.82256  & 8.2720$\pm$0.0028 & 1.407 $\pm$ 0.023 & 0.924 & B9.5V & 9\\
  HD 43519A & 93.13418 & -61.47399 & 7.1102 $\pm$0.0031 & 4.015$\pm$0.066 & 1.505 & B9.5V & 8\\
  HD 56024 & 107.95909 & -58.85062 & 8.7035$\pm$0.0028 & 2.044$\pm$0.056 & 2.594 & B9V & 11\\
  HD 62153A & 113.84019 & -74.27486 & 7.0581 $\pm$ 0.0028 &  	3.943 $\pm$ 0.029 & 1.049 &  B9IV & 5\\
  HD 198174 & 312.32344 & -25.78134 &  5.8341 $\pm$ 0.0028 & 6.666$\pm$0.187 & 2.721 & B8II & 6\\
  HD 208433 & 329.23152 & -35.35984 & 7.5530$\pm$0.0028 & 3.581$\pm$0.074 & 1.328 & B9.5V & 13\\
    \hline
    
    \end{tabular}
    \caption{Observed sample information extracted from Gaia DR3\protect\citep{gaia3} { (right ascension, declination, {\it Gaia } apparent magnitude, parallax, rectilinear unit weighted error)} and SIMBAD {\bf (spectral type)}. N is a number of SALT observations.}
    \label{tab:inp}
\end{table*}

\subsection{methods}

We use the same code, which was developed and tested to fit the FEROS spectra of HD 20784 \cite{hd20784} to analyze the new SALT spectra. It assumes SB2 configuration and extracts radial velocities $\rv$, effective temperatures $\teff$, metallicities $\feh$, microturbulences $\Vmic$, and projected rotational velocities $\vsini$ for two spectral components, performing full spectrum fitting using a neural network based spectral model. The metallicity was fixed to the same value for two components. This spectral model was trained on a synthetic model grid, generated with the GSSP code \cite{gssp}. Normalization is done simultaneously with fitting; see \cite{sarah2022} for details. 
\par
All RV measurements are collected in Tables~\ref{tab5}, \ref{tab6}.
Usually RV of brighter component in SB2 are measured with higher precision; however, in our case there are three systems where it rotates significantly faster than the dim component, thus its spectral lines are significantly broadened. For such a system, narrow lines of the secondary will allow for more precise RV measurements. Since statistical uncertainties for our RVs are around $\leq0.2~\kms$, we { artificially increased them by $0.5~\kms$ for stars with $\vsini<100~\kms$ and by $2.5~\kms$ otherwise; see Section~\ref{sec:hd208433}.}    
We use radial velocities of the component with better precision with the Generalized Lomb Scargle ({\sc GLS}) code \citep{gls} to fit the circular and Keplerian orbits.  We find that for all SB2 systems, except HD 56024, eccentricity of the Keplerian orbit is zero; so the circular orbit is a valid approximation. Therefore, we used output of {\sc GLS} to fit both RV time series simultaneously using the {\sc scipy.optimise.curve\_fit} function. Orbital parameters include systemic velocity $\gamma$, period $P$, conjunction time $t_0$,  and radial velocity semi-amplitudes for both components $K_{1,2}$. For two systems: HD 198174 and HD 208433 we didn't find any reasonable orbital solution.
\subsubsection{SB3 analysis}
For HD 56024 we were unable to get a good SB2 orbit, therefore we explored its spectra in detail. We found that it is actually a SB3 system consisting of a bright (>90\% of total light) fast rotating star and two dim slow rotators. Since narrow lined components look nearly identical, we set their parameters to be same, with the exception of $\rv$. It allows us to simplify the problem and save computational time. With these assumptions in mind, we fit all spectra and derived radial velocities and parameters for spectral components.  
\par
Based on radial velocities we conclude that HD 56024 is a hierarchical triple (1+2): a bright star orbiting an inner, close binary. Preliminary outer orbit can be derived based only on $\rv_3$ assuming it is SB1 with {\sc GLS}. Since we assumed components in the inner binary to be identical, we need to sort them out before fitting the orbit. We made it using a fit of the circular orbit for the absolute difference $|\rv_1-\rv_2|$ with {\sc GLS}, see \cite{j115} for details. Once sorting is finished, we fitted all available $\rv$ together, in a hierarchical triple configuration model, constructed using the Keplerian orbit provided in {\sc PyAstronomy} package \citep{pyasl}, using the{\sc scipy.optimise.curve\_fit} function.

\section{Results}
We collect all spectroscopic parameters in Tables~\ref{tab2},\ref{tab:two} and orbital solutions listed in Tables \ref{tab3} and \ref{tab1}.
Below we present a short summary for each individual system. 
\begin{table}[H]
\caption{Spectroscopic parameters. Here we provide mean and standard deviation among all spectra for each system.\label{tab2}}
	\begin{adjustwidth}{-\extralength}{0cm}
		\begin{tabularx}{\fulllength}{CCCCC}
			\toprule
			\textbf{System}	& \textbf{HD 20784}	& \textbf{HD 43519A}     & \textbf{HD 56024$^\ast$} & \textbf{HD 62153A}\\
			\midrule
\multirow[m]{2}{*}{$\teff$, K}	& 9479$\pm$135	&	10756$\pm$563 & 	11867$\pm$75		& 11376$\pm$261\\
			  	                   & 10364$\pm$80	&	10067$\pm$155& 7308$\pm$166			& 10771$\pm$551\\
                   \midrule
\multirow[m]{2}{*}{$\logg$, cgs} & 3.51$\pm$0.14&	3.88$\pm$0.16& 3.77$\pm$0.02			& 3.84$\pm$0.27\\
			  	                  & 3.77$\pm$0.10&	4.35$\pm$0.14 & 3.62$\pm$0.31			& 4.27$\pm$0.68\\
\midrule
$\feh^{**}$, dex &  0.01$\pm$0.01  &  -0.10$\pm$0.05 & -0.21$\pm$0.24  &  0.24$\pm$0.07 \\
\midrule
\multirow[m]{2}{*}{$\Vmic,\,\kms$}    & 1.3$\pm$0.2&	1.6$\pm$0.7 & 2.2$\pm$1.9	& 0.3$\pm$0.2\\
			  	                  & 0.9$\pm$0.2&	0.06$\pm$0.09 & 2.8$\pm$0.6			& 1.4$\pm$1.9\\
\midrule
\multirow[m]{2}{*}{$\vsini,\,\kms$}    & 30$\pm$4 &	120$\pm$6& 248$\pm$7 & 17$\pm$6$^{***}$\\
			  	                         & 17$\pm$2	&	14$\pm$7 & 11$\pm$2	& 13$\pm$7\\
            \bottomrule
		\end{tabularx}
	\end{adjustwidth}
	\noindent{\footnotesize{(*) { First row represents parameters of third component -brightest one in the triple system,} second row represents identical parameters for dim components of triple system. (**) fixed to same value for all components. (***) we excluded one measurement $\vsini=96~\kms$ taken near conjunction from the mean computation}}
\end{table}

\begin{table}[H]
\caption{Spectroscopic parameters for HD 198174 and HD 208433. Here we provide mean and standard deviation among all spectra for each system.\label{tab:two}}
	\begin{adjustwidth}{-\extralength}{0cm}
		\begin{tabularx}{\fulllength}{CCC}
			\toprule
			\textbf{System}	& \textbf{HD 198174}	& \textbf{HD 208433}\\
			\midrule
\multirow[m]{2}{*}{$\teff$, K}	& 13453$\pm$746		& 10967$\pm$37\\
			  	                   & 9686$\pm$1213		& 8013$\pm$38\\
                   \midrule
\multirow[m]{2}{*}{$\logg$, cgs} & 3.00$\pm$0.01			& 4.19$\pm$0.01\\
			  	                  & 3.27$\pm$0.38			& 4.25$\pm$0.08\\
\midrule
$\feh^{*}$, dex &  -0.32$\pm$0.03    &  -0.01$\pm$0.01 \\
\midrule
\multirow[m]{2}{*}{$\Vmic,\,\kms$}    & 0.15$\pm$0.31& 1.00$\pm$0.01\\
			  	                  & 0.0$\pm$0.1&	 1.31$\pm$0.27\\
\midrule
\multirow[m]{2}{*}{$\vsini,\,\kms$}    & 50$\pm$8  & 168$\pm$1\\
			  	                         & 44$\pm$3	& 27$\pm$1\\
            \bottomrule
		\end{tabularx}
	\end{adjustwidth}
	\noindent{\footnotesize{(*) fixed to same value for all components.}}
\end{table}

\begin{table}[H]
\caption{Circular orbits.\label{tab3}}
	\begin{adjustwidth}{-\extralength}{0cm}
		\begin{tabularx}{\fulllength}{CCCC}
			\toprule
			\textbf{System}	& \textbf{HD 20784}	& \textbf{HD 43519A}     & \textbf{HD 62153A} \\
            \midrule
$P$, days 		& 	3.58221(fixed)	  & 8.188$\pm$0.001 & 4.545$\pm$0.011\\
$t_0$, days\textsuperscript{1} &   60259.094$\pm$0.001  & 60258.198$\pm$0.004 & 60353.693$\pm$0.007\\
$\gamma,\,\kms$ &    -21.55$\pm$0.10 & 26.11$\pm$0.26 & 15.80$\pm$0.18\\
$K_{2},\,\kms$   & 74.09$\pm$0.24 & 91.85$\pm$0.45 & 44.84$\pm$0.43\\
$q=K_1/K_2=M_2/M_1$    &  0.856$\pm$ 0.004  &  0.696$\pm$0.016 &  0.605$\pm$0.011\\            
            \bottomrule
		\end{tabularx}
	\end{adjustwidth}
	\noindent{\footnotesize{\textsuperscript{1} MJD}}
\end{table}

\begin{table}[H] 
\small 
\caption{Orbital parameters for HD 56024.\label{tab1}}
\begin{tabularx}{\textwidth}{CCC}
\toprule
\textbf{Parameter}	& \textbf{HD 56024 inner orbit}	& \textbf{HD 56024 outer orbit}\\
\midrule
$P$, days 		& 	14.3834$\pm$0.0002	& 403.73$\pm$3.84 \\
$t_p$, days\textsuperscript{1}  &  60264.375$\pm$0.008  & 58854.273$\pm$16.480\\
$e$		        &  0.194$\pm$0.001  & 0.338$\pm$0.004 \\
$\omega^\circ$  &  31.38$\pm$0.19  & 305.4$\pm$1.6 \\
$\gamma,\,\kms$ &  --  & 9.93$\pm$0.06 \\
$K_{i},\,\kms$&  57.65$\pm$0.06, 58.60$\pm$0.12  & 18.47$\pm$0.12\textsuperscript{2}, 20.76$\pm$0.41 \\
\bottomrule
\end{tabularx}

\noindent{\footnotesize{\textsuperscript{1} MJD}, \textsuperscript{2} for barycenter of inner binary.}
\end{table}





\subsection{HD 20784}

\begin{figure}[H]
\begin{adjustwidth}{-\extralength}{0cm}
\centering
\includegraphics[width=12.5cm]{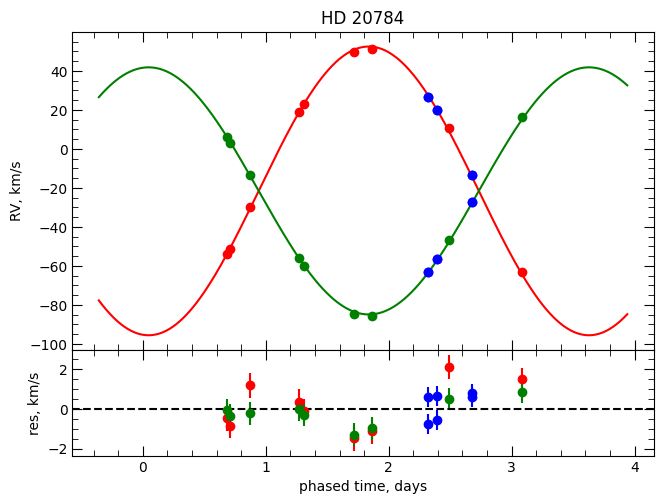}

\end{adjustwidth}
\caption{Circular orbit fit for HD 20784. { Blue circles are RV measurements from FEROS spectra \protect\citep{hd20784}}.\label{fig2a}}
\end{figure}

This is a SB2 system with relatively narrow spectral lines. We show a circular orbit fit for it in Figure~\ref{fig2a}. Fundamental parameters for HD 20784 were estimated for the first time in \cite{hd20784} using TESS \citep{tess1,tess2} photometry and RV from FEROS spectra, however all these spectra were taken near the conjunction, so these RVs were not used during the detailed modelling with {\sc PHOEBE} \citep{phoebe}. In this work we are confirming the circular orbit for this SB2 and improve estimation of projected semi-major axis $a \sin{i}=9.735\pm0.038~R_\odot$ (previous estimate $a \sin{i}=9.7^{+3.6}_{-3.2}~R_\odot$), thanks to our orbital solution.  We note that we used only new SALT based RV data to get this solution, which has a significantly smaller timebase than TESS photometry, { thus we fixed period to the value $P=3.58221$ days from \cite{hd20784}. For verification, we also plotted previous RV measurements from FEROS, and they agree very well with the new orbit. } The new value of mass ratio is slightly  smaller than previous estimate.
\par
HD 20784 has radial velocity measurement in RAVE DR6 \citep{dr6rave}: $\rv=86.92 \pm33.76~\kms$, which is not compatible with our orbital solution. Possibly, this is because the Ca triplet is poorly visible in the spectrum of the hot stars. This system is also absent in RAVE based SB1 and SB2 catalogs \citep{sb1rave2,sb2rave}. 

\subsection{HD 43519 A}

\begin{figure}[H]
\begin{adjustwidth}{-\extralength}{0cm}
\centering
\includegraphics[width=12.5cm]{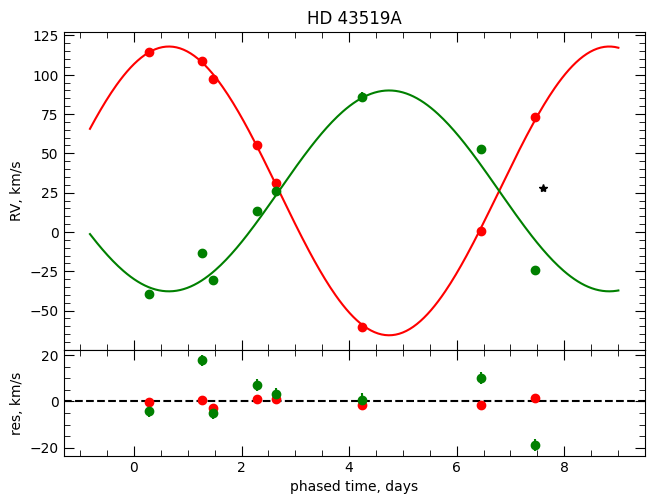}

\end{adjustwidth}
\caption{Circular orbit fit for HD 43519A.{ Black asterisk shows RV measurement from APOGEE for HD 43519B}\label{fig2b}}
\end{figure}

We show the circular orbit for this system in Figure~\ref{fig2b}. More precise RV of the secondary were fitted very well, while the primary RV shows some deviation because $\vsini_1\sim120~\kms$ of the primary is quite large. Primary is also hotter and heavier than secondary. 
\par
APOGEE DR17 \citep{apogee17} is observed nearby to ($\sim$1") the star HD 43519 B \footnote{Gaia DR3 548133010746031616} at MJD=58912.078 days and provided these parameters: $\rv=27.70\pm0.47\,\kms$, $\teff=9678$ K, $\logg=4.33$ cgs, $\feh=-1.68$ dex. Its radial velocity is very similar to the systemic velocity of the SB2 solution $\gamma=26.22\pm0.47~\kms$, so this nearby star can be bound to our SB2, forming a wide hierarchical triple. Gaia DR3 astrometry also supports this idea, see \cite{widebinary} where { authors provide chance-alignment probability of only $\sim10^{-4}$} and estimated projected separation as $a\sim271$ a.u. with an eccentricity $e=0.90^{1.00}_{0.66}$\footnote{{ given with upper and lower limits of 68\% credible interval.}} for this wide system.

\subsection{HD 56024}

\begin{figure}[H]
\begin{adjustwidth}{-\extralength}{0cm}
\centering
\subfloat[\centering]{\includegraphics[width=17.cm]{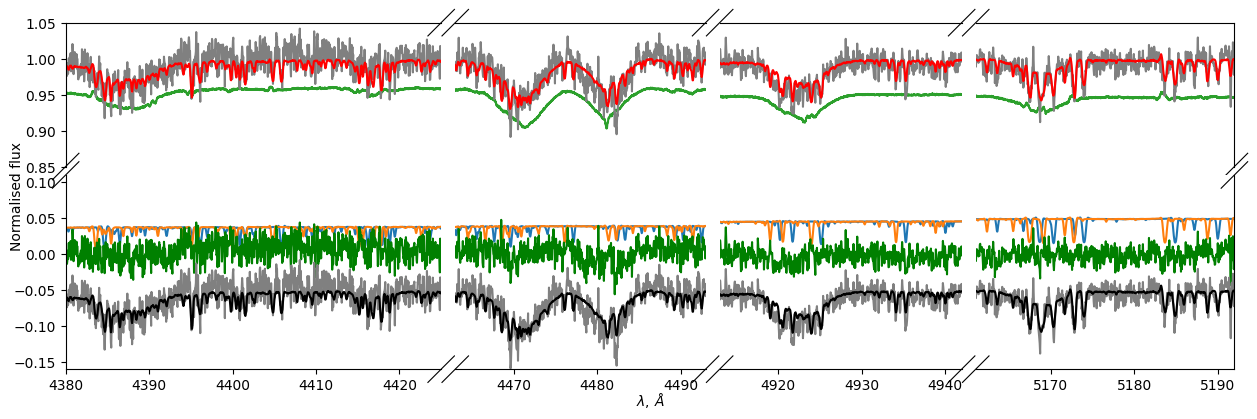}}\\
\subfloat[\centering]{\includegraphics[width=17.cm]{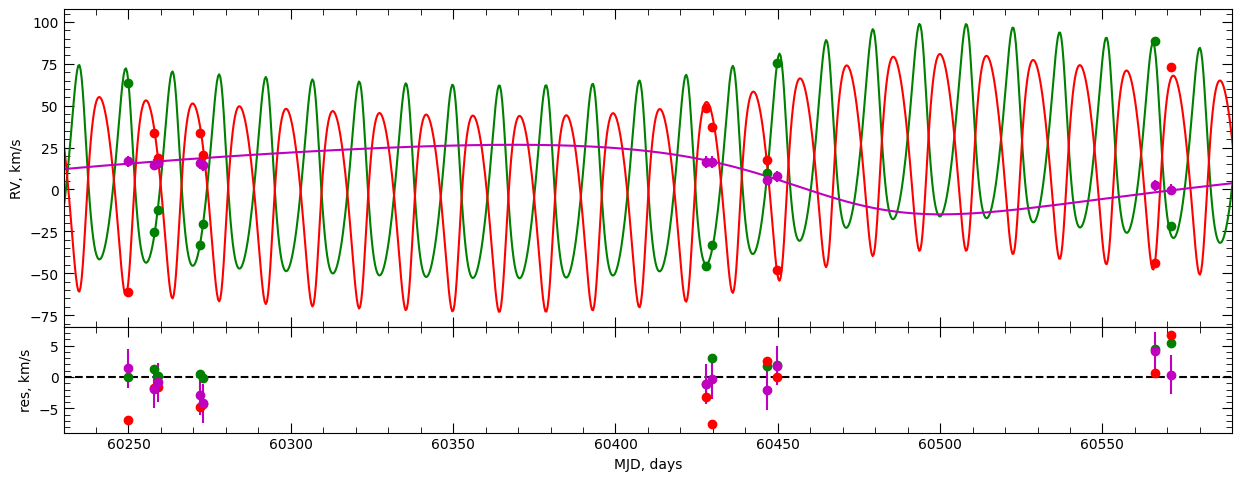}}\\
\subfloat[\centering]{\includegraphics[width=17.cm]{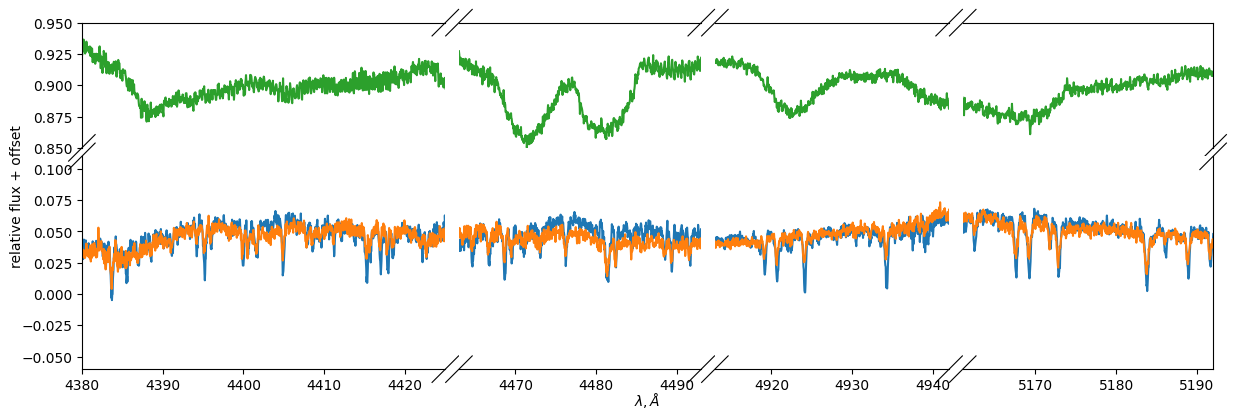}}

\end{adjustwidth}

\caption{HD 56024: {\bf (a)} Example for fitting with three spectral components in selected spectral intervals. The hot fast rotator is shown as a dark green line, while two cool components are shown as blue and orange lines respectively. Fit residuals are shown as a green line around zero. To improve visibility and readability of the Figure, we remove the gaps and plot the observed spectrum and best fit model twice: without offset (gray and red lines) and with offset -1.05 (gray and black lines) {\bf (b)} Orbit fit with hierarchical triple configuration. {\bf (c)} Spectral disentangling results by {\sc FD3} using SB3 orbit. { Here we removed sine-like modulations and shifted derived components with arbitrary offsets, to make it similar to panel (a)}\label{fig1}}
\end{figure}

This is a hierarchical triple system, which is clearly resolved in SALT spectra, see Figure~\ref{fig1} panel (a). The brightest component is a hot, fast rotating star, while the two others are significantly cooler and rotate much slower. We found that the inner binary have an eccentric orbit with period $P_{\rm in}\sim14.3$ days, while outer orbital period $P_{\rm out}\sim403.7$ days, which is $\sim28$ times larger, see panel (b). Currently SALT observations have relatively small phase coverage for the outer orbit. To check our orbital solutions for HD 56024, we fed its spectra with orbital solution to spectral disentangling code {\sc FD3} \citep{fd3}, which can handle hierarchical triple configuration. { We provided identical light factors $LF=1$ for all components, therefore disentangled spectra will the have same baseline 0.33 \citep{fakesb2} }. All three spectral components were separated and they are consistent with previous results from spectral fitting. {\sc FD3} works in Fourier space, so derived spectral components are shifted relative to their baselines with sine-like { modulations \protect\citep{fd3norm}, which we tried to remove by fitting the high-order Chebyshev polynomial, to make comparison with panel (a) easier}, see panel (c) in Figure \ref{fig1}. Also, no further optimisation of orbital parameters was required\footnote{ { we run {\sc FD3} with 10 initialisations of 1000 iterations, but found no improvement.} }, which proves that our solution is reliable. Based on radial velocity semi-amplitudes for the outer orbit, the close twin binary is slightly more massive than the hot, fast rotating primary $q_{\rm out}=M_{1-2}/M_3\equiv K_3/K_{1-2}=1.12\pm0.02$, while the mass ratio in the inner close system is near unity $q_{\rm in}=M_2/M_1\equiv K_1/K_2=0.98\pm0.02$. { Kepler's third law allow us to estimate minimal masses: $(M_1+M_2)\sin{i_{1-2}}^3\sim2.2~M_\odot$ and $(M_1+M_2+M_3)\sin{i_{3}}^3\sim2.1~M_\odot$ using the  inner and outer orbit's parameters respectively. Obviously, the outer orbit has significantly smaller inclination ($i_3<i_{1-2}$) to the sky-plane, than the inner one.}  
\par
APOGEE DR17 \citep{apogee17} observed this star twice: on MJD=58503.293, 58831.359 days and provided { the following} parameters assuming a single star: $\rv=17.91\pm0.17, -10.64\pm0.44\,\kms$, $\teff=7663, 10856$ K, $\logg=4.25, 4.11$ cgs, $\feh=-2.21, -1.07$ dex. It is highly likely their pipeline completely removed bright, fast rotator features during the normalisation of the first spectrum (see \citep{ttau}, where a similar problem was described), so they get parameters for one of the cool components. Note that the first spectrum indicates an extremely metal-poor star, since relative contribution of the cool components is small; while in the second spectrum derived parameters corresponds to a bright, hot primary.

\subsection{HD 62153A}
\begin{figure}[H]
\begin{adjustwidth}{-\extralength}{0cm}
\centering
\includegraphics[width=12.5cm]{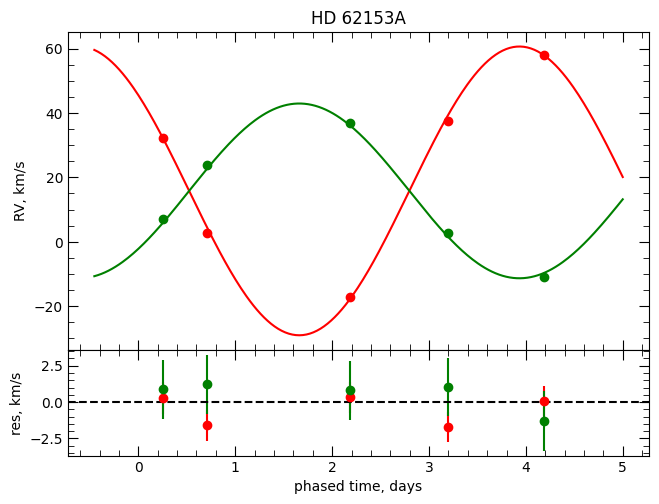}

\end{adjustwidth}
\caption{Circular orbit fit for HD 62153A.\label{fig2c}}
\end{figure}

The orbital solution for this system is presented in Figure~\ref{fig2c}. Both components exhibit relatively narrow spectral lines. The primary star is hotter and more massive than the secondary.

The derived orbital period, $P = 4.545 \pm 0.011$~d, is in good agreement with the rotational period, $P_{\rm rot} = \nu_{\rm rot}^{-1} = 1/0.216367 = 4.622$~d, measured by \cite{balona}, who inferred the presence of photospheric spots on one or both stars in this system. The system was first identified as SB2 by \cite{1983A&AS...52..471A}, who determined a mass ratio of $q = 0.58 \pm 0.02$, in good agreement with our estimate of $q = 0.605 \pm 0.011$.

\subsection{HD 198174}

\begin{figure}[H]
\begin{adjustwidth}{-\extralength}{0cm}
\centering
\includegraphics[width=12.5cm]{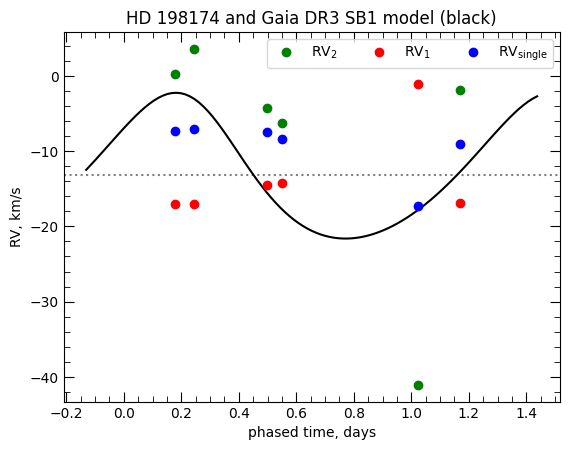}
\end{adjustwidth}
\caption{Gaia DR3 SB1 orbit and RV from binary and single-star model for HD 198174. Dotted line shows systemic velocity. \label{fig2d}}
\end{figure}

{ HD 198174 is the sole bright giant (its SIMBAD spectral type is B9\,II) in our sample, and it is the only object that appears in the \textit{Gaia} Non-Single Star catalogue \citep{gaia_dr3multiple} with a SB1 orbital solution\footnote{period $P=1.3071\pm0.0001$ d, center of mass velocity $\gamma=-13.20\pm0.41~\kms$, periastron longitude $\omega=30.28\pm20.34$ deg, semi-amplitude $K=9.68\pm0.53~\kms$, eccentricity $e=0.15\pm0.06$, periastron passage time (MJD) $t_0= 55196.48139\pm0.09287$ d}. Unfortunately, our $\rv$ measurements are inconsistent with this orbital solution for both components (see Figure~\ref{fig2d}); however we should mention that HD 198174 is the only system where none of the spectra were showing clear composite structure. Therefore, we also included single-star model $\rv$ measurements in the Figure~\ref{fig2d}. We have only one good match with SB1 orbit, although it looks like accidental agreement. The Renormalised Unit Weight Error (RUWE) for this system is relatively high, with a value of 2.72. This suggests that the source is unlikely to be a single star. Consequently, forthcoming \textit{Gaia} data releases may yield an improved orbital solution for this object.   }

\subsection{HD 208433}
\label{sec:hd208433}

\begin{figure}[H]
\begin{adjustwidth}{-\extralength}{0cm}
\centering
\includegraphics[width=12.5cm]{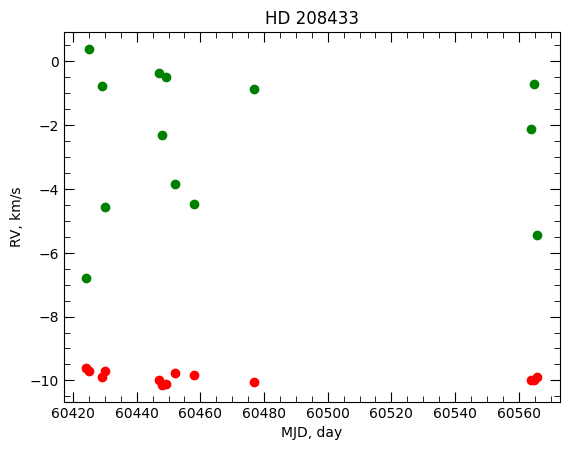}
\end{adjustwidth}
\caption{ RV data for HD 208433.\label{fig2e}}
\end{figure}

{ Although we obtained 13 radial‑velocity measurements for this system, we did not detect any statistically significant temporal variability that would support its classification as SB2; see Figure~\ref{fig2e}. The narrow-lined ($\vsini = 28~\kms$) spectral component exhibits an approximately constant radial velocity of $\rv_2 = -9.90 \pm 0.16~\kms$, whereas the bright, rapidly rotating component ($\vsini = 168~\kms$) shows only marginal variations in radial velocity, with $\rv_1 = -2.5 \pm 2.5~\kms$. A {\sc GLS} periodogram analysis did not reveal any statistically significant periodic signal in $\rv_1$, and we therefore interpret the observed variations as arising from random measurement errors. Consequently, we propose that HD 208433 is a chance alignment of two unrelated stars, a scenario further supported by the relatively low RUWE value of 1.328 reported in \textit{Gaia} DR3. Under this assumption of constant $\rv$ for both spectral components, we use these measurements to characterize the performance of our radial-velocity determination method for rapidly rotating stars ($\sigma\rv = 2.5~\kms$) and slowly rotating stars ($\sigma\rv = 0.16~\kms$). }

\section{Discussion}

\par
The discovery of the hierarchical nature of SB3 in the HD 56024 system opens a great opportunity to study the formation scenario and dynamical evolution of this multiple system. This system has many high quality TESS observations (currently 44 sectors are available) which show low amplitude ($\leq0.1\%$) sine like variability with a period of $\sim1.2$ days. \cite{sarah2022} classified it as a slowly pulsating B-type variable, which can be a hot, fast rotating primary component; however, this can also be  explained by spots on one or both cool components of the inner binary.  We think that thanks to the relatively large outer orbit we can identify which of the  SB3 components is the source of this variability through the light time travel effect (LTTE)\citep{ttau}. The expected semi-amplitude for the ``light" orbit is $A=a_{\rm out} \sin{i_{\rm out}}/c\sim0.004$ days, with $c$ - speed of light, which should be detectable in TESS data. 
Also, ``light" orbit will help us to constrain the value of the outer period and test if it is in resonance 1:28 with the inner period. { Additional SALT observations are being planned for HD~56024 (Kovalev et al, in prep)}.   
\par
In this study we used only data from SALT spectra; however, in principle it is possible to combine our analysis with previous observations from \cite{1983A&AS...52..471A,apogee17,sarah2022,dr6rave}.
\section{Conclusions}

We confirmed four spectroscopic binary candidates using new observations by SALT. Three SB2 systems have circular orbits with periods less than 10 days, while one hierarchical triple has a close binary with an inner elliptical orbit with period $\sim14$ days, consisting of nearly identical components, and a fast rotator on an outer elliptical orbit with period $\sim400$ days. For two other SB2 candidates, our observations did not { allow orbit determination}. 



\vspace{6pt} 





\authorcontributions{
Conceptualization, M.K. ; methodology, M.K.; software, M.K. and A.K.; validation, M.K., A.K. and O.M.; formal analysis, M.K.; investigation, M.K.; resources, A.K.; data curation, A.K.; writing---original draft preparation, M.K.; writing---review and editing, M.K., A.K. and O.M. ; visualization, M.K.; supervision, O.M.; project administration, A.K.; funding acquisition, A.K. All authors have read and agreed to the published version of the manuscript.}

\funding{This research received no external funding}

\dataavailability{SALT spectra will be publicly available on \url{ssda.saao.ac.za} }

\acknowledgments{ {\bf We are grateful to the anonymous referee for a constructive report. We thank Oliver Guy for a careful proof-reading of the manuscript. } Spectral observations reported in this paper were obtained with the Southern African Large Telescope (SALT) program 2023-2-MLT-001 (PI: Kniazev). A.\,K. acknowledges support from the National Research Foundation (NRF) of South Africa. O.~M. thanks the Foundation for the Advancement of Theoretical Physics and Mathematics ``BASIS'' for its support. The study was conducted under the state assignment of Lomonosov Moscow State University. A part of the work was carried out under the state assignment of the Institute of Astronomy of RAS. This work is also supported by International Centre of Supernovae (ICESUN), Yunnan Key Laboratory of Supernova Research (No. 202505AV340004)}

\conflictsofinterest{The authors declare no conflicts of interest.}



\abbreviations{Abbreviations}{
The following abbreviations are used in this manuscript:
\\
\noindent 
\begin{tabular}{@{}ll}
APOGEE & Apache point observatory Galaxy evolution experiment\\
RAVE & RAdial Velocity Experiment\\
$\rv$ & radial velocity\\
LTTE & light time travel effect\\
TESS & Transiting exoplanet survey satelite\\
GLS & Generalised Lomb-Scargle\\
SALT & South African Large Telescope\\
SB1 & single line spectroscopic binary\\
SB2 & double lined spectroscopic binary\\
SB3 & spectroscopic triple\\
ST & spectral type from SIMBAD\citep{simbad} \\
RUWE & Rectilinear Unit Weight Error in \textit{Gaia} DR3\citep{gaia3}\\
DR & data release
\end{tabular}
}

\appendixtitles{no} 
\appendixstart
\appendix
\section[\appendixname~\thesection]{RV data}

\begin{table}[H] 
\caption{Radial velocity measurements. Statistical uncertainties are nominal, see text for details.\label{tab5}}
\begin{tabularx}{\textwidth}{CCCC}
\toprule
\textbf{MJD, day}	& \textbf{$\rv_2,\,\kms$}	& \textbf{$\rv_1,\,\kms$} & $\rv_3,\,\kms$\\
\midrule
HD 20784 \\
60257.047 & -54.16 $\pm$ 0.11 & 5.87 $\pm$ 0.06 & \\
60257.073 & -51.39 $\pm$ 0.13 & 2.92 $\pm$ 0.07 & \\
60258.081 & 49.49 $\pm$ 0.13 & -84.92 $\pm$ 0.07 & \\
60260.812 & -29.87 $\pm$ 0.12 & -13.64 $\pm$ 0.08 & \\
60261.806 & 51.35 $\pm$ 0.15 & -85.87 $\pm$ 0.08 & \\
60263.031 & -63.25 $\pm$ 0.08 & 16.23 $\pm$ 0.05 & \\
60264.793 & 18.86 $\pm$ 0.11 & -55.85 $\pm$ 0.07 & \\
60264.834 & 22.82 $\pm$ 0.09 & -59.93 $\pm$ 0.06 & \\
60266.018 & 10.81 $\pm$ 0.10 & -46.94 $\pm$ 0.06 & \\
\midrule
HD 43519 A\\
60249.951 & 31.23 $\pm$ 0.05 & 26.22 $\pm$ 0.08 & \\
60256.973 & 97.35 $\pm$ 0.06 & -30.34 $\pm$ 0.06 & \\
60261.949 & 0.50 $\pm$ 0.08 & 52.62 $\pm$ 0.07 & \\
60262.966 & 72.89 $\pm$ 0.06 & -24.22 $\pm$ 0.08 & \\
60265.975 & 55.39 $\pm$ 0.03 & 13.24 $\pm$ 0.05 & \\
60267.928 & -60.35 $\pm$ 0.54 & 85.84 $\pm$ 0.42 & \\
60427.736 & 114.36 $\pm$ 0.24 & -39.39 $\pm$ 0.20 & \\
60428.714 & 108.55 $\pm$ 0.04 & -13.14 $\pm$ 0.07 & \\
\midrule
HD 56024\\
60250.001 & 63.65 $\pm$ 0.10 & -61.28 $\pm$ 0.10 & 16.81 $\pm$ 0.42 \\
60258.011 & -25.46 $\pm$ 0.11 & 33.41 $\pm$ 0.11 & 14.77 $\pm$ 0.42 \\
60259.005 & -12.12 $\pm$ 0.08 & 18.66 $\pm$ 0.09 & 15.93 $\pm$ 0.31 \\
60271.941 & -33.22 $\pm$ 0.09 & 33.96 $\pm$ 0.09 & 15.64 $\pm$ 0.35 \\
60272.989 & -20.47 $\pm$ 0.14 & 20.73 $\pm$ 0.14 & 14.54 $\pm$ 0.55 \\
60427.773 & -45.75 $\pm$ 0.16 & 48.76 $\pm$ 0.17 & 16.35 $\pm$ 0.60 \\
60429.762 & -32.91 $\pm$ 0.13 & 37.03 $\pm$ 0.13 & 16.25 $\pm$ 0.54 \\
60446.692 & 10.06 $\pm$ 0.11 & 17.77 $\pm$ 0.11 & 5.73 $\pm$ 0.32 \\
60449.695 & 75.69 $\pm$ 0.12 & -48.03 $\pm$ 0.12 & 7.79 $\pm$ 0.44 \\
60566.130 & 88.42 $\pm$ 0.10 & -43.82 $\pm$ 0.10 & 2.42 $\pm$ 0.33 \\
60571.125 & -21.88 $\pm$ 0.10 & 73.11 $\pm$ 0.10 & -0.14 $\pm$ 0.38 \\
\midrule
HD 62153 A\\
60348.882 & 32.13 $\pm$ 0.08 & 6.95 $\pm$ 0.02 & \\
60350.813 & -17.24 $\pm$ 0.07 & 36.79 $\pm$ 0.02 & \\
60351.819 & 37.58 $\pm$ 0.04 & 2.61 $\pm$ 0.02 & \\
60352.814 & 57.87 $\pm$ 0.05 & -10.94 $\pm$ 0.06 & \\
60353.881 & 2.68 $\pm$ 0.06 & 23.97 $\pm$ 0.06 & \\

\bottomrule
\end{tabularx}
\end{table}

\begin{table}[H] 
\caption{Radial velocity measurements for HD 198174 and HD 208433. Statistical uncertainties are nominal, see text for details.\label{tab6}}
\begin{tabularx}{\textwidth}{CCCC}
\toprule
\textbf{MJD, day}	& \textbf{$\rv_{\rm single},\,\kms$}	& \textbf{$\rv_2,\,\kms$} & $\rv_1,\,\kms$\\
\midrule
HD 198174\\
60257.807 & -7.45 $\pm$ 0.10 & -4.24 $\pm$ 0.10 & -14.52 $\pm$ 0.08  \\
60258.795 & -7.36 $\pm$ 0.08 & 0.17 $\pm$ 0.08 & -16.98 $\pm$ 0.07  \\
60259.787 & -9.12 $\pm$ 0.08 & -1.95 $\pm$ 0.07 & -16.86 $\pm$ 0.06  \\
60261.783 & -8.44 $\pm$ 0.08 & -6.22 $\pm$ 0.07 & -14.23 $\pm$ 0.10  \\
60262.783 & -7.09 $\pm$ 0.10 & 3.57 $\pm$ 0.10 & -16.99 $\pm$ 0.07  \\
60419.108 & -17.25 $\pm$ 0.16 & -41.04 $\pm$ 0.16 & -1.08 $\pm$ 0.13  \\
\midrule
HD 208433\\
60424.152 & & -9.62 $\pm$ 0.13 & -6.79 $\pm$ 0.06  \\
60425.130 & & -9.69 $\pm$ 0.13 & 0.38 $\pm$ 0.06  \\
60429.121 & & -9.89 $\pm$ 0.09 & -0.80 $\pm$ 0.05  \\
60430.113 & & -9.70 $\pm$ 0.12 & -4.58 $\pm$ 0.06  \\
60447.085 & & -10.00 $\pm$ 0.06 & -0.38 $\pm$ 0.03  \\
60448.068 & & -10.14 $\pm$ 0.07 & -2.32 $\pm$ 0.04  \\
60449.070 & & -10.12 $\pm$ 0.08 & -0.49 $\pm$ 0.04  \\
60452.050 & & -9.77 $\pm$ 0.09 & -3.84 $\pm$ 0.05  \\
60458.039 & & -9.83 $\pm$ 0.10 & -4.49 $\pm$ 0.05  \\
60476.996 & & -10.03 $\pm$ 0.09 & -0.87 $\pm$ 0.05  \\
60563.754 & & -10.00 $\pm$ 0.11 & -2.12 $\pm$ 0.05  \\
60564.773 & & -10.00 $\pm$ 0.12 & -0.71 $\pm$ 0.07  \\
60565.768 & & -9.90 $\pm$ 0.09 & -5.44 $\pm$ 0.05  \\

\bottomrule
\end{tabularx}
\end{table}


\begin{adjustwidth}{-\extralength}{0cm}

\reftitle{References}



\begin{thebibliography}{999}

\bibitem[Moe and Di~Stefano(2017)]{moe2017mind}
Moe, M.; Di~Stefano, R.
\newblock Mind your Ps and Qs: the interrelation between period (P) and mass-ratio (Q) distributions of binary stars.
\newblock {\em The Astrophysical Journal Supplement Series} {\bf 2017}, {\em 230},~15.

\bibitem[Han et~al.(2020)Han, Ge, Chen, and Chen]{han2020binary}
Han, Z.W.; Ge, H.W.; Chen, X.F.; Chen, H.L.
\newblock Binary population synthesis.
\newblock {\em Research in Astronomy and Astrophysics} {\bf 2020}, {\em 20},~161.

\bibitem[{Gaia Collaboration} et~al.(2022){Gaia Collaboration}, {Arenou}, {Babusiaux}, {Barstow}, {Faigler}, {Jorissen}, {Kervella}, {Mazeh}, {Mowlavi}, {Panuzzo}, {Sahlmann}, {Shahaf}, {Sozzetti}, {Bauchet}, {Damerdji}, {Gavras}, {Giacobbe}, {Gosset}, {Halbwachs}, {Holl}, {Lattanzi}, {Leclerc}, {Morel}, {Pourbaix}, {Re Fiorentin}, {Sadowski}, {S{\'e}gransan}, {Siopis}, {Teyssier}, {Zwitter}, {Planquart}, {Brown}, {Vallenari}, {Prusti}, {de Bruijne}, {Biermann}, {Creevey}, {Ducourant}, {Evans}, {Eyer}, {Guerra}, {Hutton}, {Jordi}, {Klioner}, {Lammers}, {Lindegren}, {Luri}, {Mignard}, {Panem}, {Randich}, {Sartoretti}, {Soubiran}, {Tanga}, {Walton}, {Bailer-Jones}, {Bastian}, {Drimmel}, {Jansen}, {Katz}, {van Leeuwen}, {Bakker}, {Cacciari}, {Casta{\~n}eda}, {De Angeli}, {Fabricius}, {Fouesneau}, {Fr{\'e}mat}, {Galluccio}, {Guerrier}, {Heiter}, {Masana}, {Messineo}, {Nicolas}, {Nienartowicz}, {Pailler}, {Riclet}, {Roux}, {Seabroke}, {Sordo}, {Th{\'e}venin}, {Gracia-Abril}, {Portell}, {Altmann}, {Andrae},
  {Audard}, {Bellas-Velidis}, {Benson}, {Berthier}, {Blomme}, {Burgess}, {Busonero}, {Busso}, {C{\'a}novas}, {Carry}, {Cellino}, {Cheek}, {Clementini}, {Davidson}, {de Teodoro}, {Nu{\~n}ez Campos}, {Delchambre}, {Dell'Oro}, {Esquej}, {Fern{\'a}ndez-Hern{\'a}ndez}, {Fraile}, {Garabato}, {Garc{\'\i}a-Lario}, {Haigron}, {Hambly}, {Harrison}, {Hern{\'a}ndez}, {Hestroffer}, {Hodgkin}, {Jan{\ss}en}, {Jevardat de Fombelle}, {Jordan}, {Krone-Martins}, {Lanzafame}, {L{\"o}ffler}, {Marchal}, {Marrese}, {Moitinho}, {Muinonen}, {Osborne}, {Pancino}, {Pauwels}, {Recio-Blanco}, {Reyl{\'e}}, {Riello}, {Rimoldini}, {Roegiers}, {Rybizki}, {Sarro}, {Smith}, {Utrilla}, {van Leeuwen}, {Abbas}, {{\'A}brah{\'a}m}, {Abreu Aramburu}, {Aerts}, {Aguado}, {Ajaj}, {Aldea-Montero}, {Altavilla}, {{\'A}lvarez}, {Alves}, {Anders}, {Anderson}, {Anglada Varela}, {Antoja}, {Baines}, {Baker}, {Balaguer-N{\'u}{\~n}ez}, {Balbinot}, {Balog}, {Barache}, {Barbato}, {Barros}, {Bartolom{\'e}}, {Bassilana}, {Becciani}, {Bellazzini}, {Berihuete},
  {Bernet}, {Bertone}, {Bianchi}, {Binnenfeld}, {Blanco-Cuaresma}, {Blazere}, {Boch}, {Bombrun}, {Bossini}, {Bouquillon}, {Bragaglia}, {Bramante}, {Breedt}, {Bressan}, {Brouillet}, {Brugaletta}, {Bucciarelli}, {Burlacu}, {Butkevich}, {Buzzi}, {Caffau}, {Cancelliere}, {Cantat-Gaudin}, {Carballo}, {Carlucci}, {Carnerero}, {Carrasco}, {Casamiquela}, {Castellani}, {Castro-Ginard}, {Chaoul}, {Charlot}, {Chemin}, {Chiaramida}, {Chiavassa}, {Chornay}, {Comoretto}, {Contursi}, {Cooper}, {Cornez}, {Cowell}, {Crifo}, {Cropper}, {Crosta}, {Crowley}, {Dafonte}, {Dapergolas}, {David}, {de Laverny}, {De Luise}, {De March}, {De Ridder}, {de Souza}, {de Torres}, {del Peloso}, {del Pozo}, {Delbo}, {Delgado}, {Delisle}, {Demouchy}, {Dharmawardena}, {Diakite}, {Diener}, {Distefano}, {Dolding}, {Enke}, {Fabre}, {Fabrizio}, {Fedorets}, {Fernique}, {Figueras}, {Fournier}, {Fouron}, {Fragkoudi}, {Gai}, {Garcia-Gutierrez}, {Garcia-Reinaldos}, {Garc{\'\i}a-Torres}, {Garofalo}, {Gavel}, {Gerlach}, {Geyer}, {Gilmore}, {Girona},
  {Giuffrida}, {Gomel}, {Gomez}, {Gonz{\'a}lez-N{\'u}{\~n}ez}, {Gonz{\'a}lez-Santamar{\'\i}a}, {Gonz{\'a}lez-Vidal}, {Granvik}, {Guillout}, {Guiraud}, {Guti{\'e}rrez-S{\'a}nchez}, {Guy}, {Hatzidimitriou}, {Hauser}, {Haywood}, {Helmer}, {Helmi}, {Sarmiento}, {Hidalgo}, {H{\l}adczuk}, {Hobbs}, {Holland}, {Huckle}, {Jardine}, {Jasniewicz}, {Jean-Antoine Piccolo}, {Jim{\'e}nez-Arranz}, {Juaristi Campillo}, {Julbe}, {Karbevska}, {Khanna}, {Kordopatis}, {Korn}, {K{\'o}sp{\'a}l}, {Kostrzewa-Rutkowska}, {Kruszy{\'n}ska}, {Kun}, {Laizeau}, {Lambert}, {Lanza}, {Lasne}, {Le Campion}, {Lebreton}, {Lebzelter}, {Leccia}, {Lecoeur-Taibi}, {Liao}, {Licata}, {Lindstr{\o}m}, {Lister}, {Livanou}, {Lobel}, {Lorca}, {Loup}, {Madrero Pardo}, {Magdaleno Romeo}, {Managau}, {Mann}, {Manteiga}, {Marchant}, {Marconi}, {Marcos}, {Marcos Santos}, {Mar{\'\i}n Pina}, {Marinoni}, {Marocco}, {Marshall}, {Polo}, {Mart{\'\i}n-Fleitas}, {Marton}, {Mary}, {Masip}, {Massari}, {Mastrobuono-Battisti}, {McMillan}, {Messina}, {Michalik}, {Millar},
  {Mints}, {Molina}, {Molinaro}, {Moln{\'a}r}, {Monari}, {Mongui{\'o}}, {Montegriffo}, {Montero}, {Mor}, {Mora}, {Morbidelli}, {Morris}, {Muraveva}, {Murphy}, {Musella}, {Nagy}, {Noval}, {Oca{\~n}a}, {Ogden}, {Ordenovic}, {Osinde}, {Pagani}, {Pagano}, {Palaversa}, {Palicio}, {Pallas-Quintela}, {Panahi}, {Payne-Wardenaar}, {Pe{\~n}alosa Esteller}, {Penttil{\"a}}, {Pichon}, {Piersimoni}, {Pineau}, {Plachy}, {Plum}, {Poggio}, {Pr{\v{s}}a}, {Pulone}, {Racero}, {Ragaini}, {Rainer}, {Raiteri}, {Ramos}, {Ramos-Lerate}, {Regibo}, {Richards}, {Rios Diaz}, {Ripepi}, {Riva}, {Rix}, {Rixon}, {Robichon}, {Robin}, {Robin}, {Roelens}, {Rogues}, {Rohrbasser}, {Romero-G{\'o}mez}, {Rowell}, {Royer}, {Ruz Mieres}, {Rybicki}, {S{\'a}ez N{\'u}{\~n}ez}, {Sagrist{\`a} Sell{\'e}s}, {Salguero}, {Samaras}, {Sanchez Gimenez}, {Sanna}, {Santove{\~n}a}, {Sarasso}, {Schultheis}, {Sciacca}, {Segol}, {Segovia}, {Semeux}, {Siddiqui}, {Siebert}, {Siltala}, {Silvelo}, {Slezak}, {Slezak}, {Smart}, {Snaith}, {Solano}, {Solitro}, {Souami},
  {Souchay}, {Spagna}, {Spina}, {Spoto}, {Steele}, {Steidelm{\"u}ller}, {Stephenson}, {S{\"u}veges}, {Surdej}, {Szabados}, {Szegedi-Elek}, {Taris}, {Taylor}, {Teixeira}, {Tolomei}, {Tonello}, {Torra}, {Torra}, {Torralba Elipe}, {Trabucchi}, {Tsounis}, {Turon}, {Ulla}, {Unger}, {Vaillant}, {van Dillen}, {van Reeven}, {Vanel}, {Vecchiato}, {Viala}, {Vicente}, {Voutsinas}, {Weiler}, {Wevers}, {Wyrzykowski}, {Yoldas}, {Yvard}, {Zhao}, {Zorec}, and {Zucker}]{gaia_dr3multiple}
{Gaia Collaboration}.; {Arenou}, F.; {Babusiaux}, C.; {Barstow}, M.A.; {Faigler}, S.; {Jorissen}, A.; {Kervella}, P.; {Mazeh}, T.; {Mowlavi}, N.; {Panuzzo}, P.;  et~al.
\newblock {Gaia Data Release 3: Stellar multiplicity, a teaser for the hidden treasure}.
\newblock {\em arXiv e-prints} {\bf 2022}, p. arXiv:2206.05595,  \href{http://arxiv.org/abs/2206.05595}{{\normalfont [arXiv:astro-ph.SR/2206.05595]}}.

\bibitem[{Tokovinin}(2025)]{tokovinin25}
{Tokovinin}, A.
\newblock {Orbits of Six Triple Systems}.
\newblock {\em \aj} {\bf 2025}, {\em 169},~124,  \href{http://arxiv.org/abs/2501.04807}{{\normalfont [arXiv:astro-ph.SR/2501.04807]}}.
\newblock {\url{https://doi.org/10.3847/1538-3881/ada3c6}}.

\bibitem[{Tanikawa} et~al.(2026){Tanikawa}, {Tajitsu}, {Honda}, {Maehara}, {Sato}, {Masuda}, {Omiya}, and {Izumiura}]{sb3japan}
{Tanikawa}, A.; {Tajitsu}, A.; {Honda}, S.; {Maehara}, H.; {Sato}, B.; {Masuda}, K.; {Omiya}, M.; {Izumiura}, H.
\newblock {Discovery of a compact hierarchical triple main-sequence star system while searching for binary stars with compact objects}.
\newblock {\em arXiv e-prints} {\bf 2026}, p. arXiv:2601.21125,  \href{http://arxiv.org/abs/2601.21125}{{\normalfont [arXiv:astro-ph.SR/2601.21125]}}.
\newblock {\url{https://doi.org/10.48550/arXiv.2601.21125}}.

\bibitem[{Antognini} and {Thompson}(2016)]{sb3sim}
{Antognini}, J.M.O.; {Thompson}, T.A.
\newblock {Dynamical formation and scattering of hierarchical triples: cross-sections, Kozai-Lidov oscillations, and collisions}.
\newblock {\em \mnras} {\bf 2016}, {\em 456},~4219--4246,  \href{http://arxiv.org/abs/1507.03593}{{\normalfont [arXiv:astro-ph.SR/1507.03593]}}.
\newblock {\url{https://doi.org/10.1093/mnras/stv2938}}.

\bibitem[{Tokovinin}(2026)]{tokovinin2}
{Tokovinin}, A.
\newblock {Mutual Orbit Alignment in Resolved Triple Systems}.
\newblock {\em \apj} {\bf 2026}, {\em 998},~151,  \href{http://arxiv.org/abs/2601.05006}{{\normalfont [arXiv:astro-ph.SR/2601.05006]}}.
\newblock {\url{https://doi.org/10.3847/1538-4357/ae3682}}.

\bibitem[{Gebruers} et~al.(2022){Gebruers}, {Tkachenko}, {Bowman}, {Van Reeth}, {Burssens}, {IJspeert}, {Mahy}, {Straumit}, {Xiang}, {Rix}, and {Aerts}]{sarah2022}
{Gebruers}, S.; {Tkachenko}, A.; {Bowman}, D.M.; {Van Reeth}, T.; {Burssens}, S.; {IJspeert}, L.; {Mahy}, L.; {Straumit}, I.; {Xiang}, M.; {Rix}, H.W.;  et~al.
\newblock {Analysis of high-resolution FEROS spectroscopy for a sample of variable B-type stars assembled from TESS photometry}.
\newblock {\em \aap} {\bf 2022}, {\em 665},~A36,  \href{http://arxiv.org/abs/2206.11280}{{\normalfont [arXiv:astro-ph.SR/2206.11280]}}.
\newblock {\url{https://doi.org/10.1051/0004-6361/202243839}}.

\bibitem[{Kaufer} et~al.(1999){Kaufer}, {Stahl}, {Tubbesing}, {N{\o}rregaard}, {Avila}, {Francois}, {Pasquini}, and {Pizzella}]{feros}
{Kaufer}, A.; {Stahl}, O.; {Tubbesing}, S.; {N{\o}rregaard}, P.; {Avila}, G.; {Francois}, P.; {Pasquini}, L.; {Pizzella}, A.
\newblock {Commissioning FEROS, the new high-resolution spectrograph at La-Silla.}
\newblock {\em The Messenger} {\bf 1999}, {\em 95},~8--12.

\bibitem[{Ricker} et~al.(2015){Ricker}, {Winn}, {Vanderspek}, {Latham}, {Bakos}, {Bean}, {Berta-Thompson}, {Brown}, {Buchhave}, {Butler}, {Butler}, {Chaplin}, {Charbonneau}, {Christensen-Dalsgaard}, {Clampin}, {Deming}, {Doty}, {De Lee}, {Dressing}, {Dunham}, {Endl}, {Fressin}, {Ge}, {Henning}, {Holman}, {Howard}, {Ida}, {Jenkins}, {Jernigan}, {Johnson}, {Kaltenegger}, {Kawai}, {Kjeldsen}, {Laughlin}, {Levine}, {Lin}, {Lissauer}, {MacQueen}, {Marcy}, {McCullough}, {Morton}, {Narita}, {Paegert}, {Palle}, {Pepe}, {Pepper}, {Quirrenbach}, {Rinehart}, {Sasselov}, {Sato}, {Seager}, {Sozzetti}, {Stassun}, {Sullivan}, {Szentgyorgyi}, {Torres}, {Udry}, and {Villasenor}]{tess}
{Ricker}, G.R.; {Winn}, J.N.; {Vanderspek}, R.; {Latham}, D.W.; {Bakos}, G.{\'A}.; {Bean}, J.L.; {Berta-Thompson}, Z.K.; {Brown}, T.M.; {Buchhave}, L.; {Butler}, N.R.;  et~al.
\newblock {Transiting Exoplanet Survey Satellite (TESS)}.
\newblock {\em Journal of Astronomical Telescopes, Instruments, and Systems} {\bf 2015}, {\em 1},~014003.
\newblock {\url{https://doi.org/10.1117/1.JATIS.1.1.014003}}.

\bibitem[{Kovalev} and {Straumit}(2023)]{hd20784}
{Kovalev}, M.; {Straumit}, I.
\newblock {Application of the binary spectral model to high-resolution spectra. First estimation of the fundamental parameters for HD 20784}.
\newblock {\em \mnras} {\bf 2023}, {\em 523},~3741--3748,  \href{http://arxiv.org/abs/2210.00863}{{\normalfont [arXiv:astro-ph.SR/2210.00863]}}.
\newblock {\url{https://doi.org/10.1093/mnras/stad1667}}.

\bibitem[{Buckley} et~al.(2006){Buckley}, {Swart}, and {Meiring}]{2006SPIE.6267E..0ZB}
{Buckley}, D.A.H.; {Swart}, G.P.; {Meiring}, J.G., {Completion and commissioning of the Southern African Large Telescope}.
\newblock In {\em {Ground-based and Airborne Telescopes. Edited by Stepp, Larry M.. Proceedings of the SPIE, Volume 6267, id. 62670Z (2006).}};  2006; Vol. 6267, {\em {Society of Photo-Optical Instrumentation Engineers (SPIE) Conference Series}}, p. 62670Z.
\newblock {\url{https://doi.org/10.1117/12.673750}}.

\bibitem[{O'Donoghue} et~al.(2006){O'Donoghue}, {Buckley}, {Balona}, {Bester}, {Botha}, {Brink}, {Carter}, {Charles}, {Christians}, {Ebrahim}, {Emmerich}, {Esterhuyse}, {Evans}, {Fourie}, {Fourie}, {Gajjar}, {Gordon}, {Gumede}, {de Kock}, {Koeslag}, {Koorts}, {Kriel}, {Marang}, {Meiring}, {Menzies}, {Menzies}, {Metcalfe}, {Meyer}, {Nel}, {O'Connor}, {Osman}, {Du Plessis}, {Rall}, {Riddick}, {Romero-Colmenero}, {Potter}, {Sass}, {Schalekamp}, {Sessions}, {Siyengo}, {Sopela}, {Steyn}, {Stoffels}, {Scholtz}, {Swart}, {Swat}, {Swiegers}, {Tiheli}, {Vaisanen}, {Whittaker}, and {van Wyk}]{2006MNRAS.372..151O}
{O'Donoghue}, D.; {Buckley}, D.A.H.; {Balona}, L.A.; {Bester}, D.; {Botha}, L.; {Brink}, J.; {Carter}, D.B.; {Charles}, P.A.; {Christians}, A.; {Ebrahim}, F.;  et~al.
\newblock {First science with the Southern African Large Telescope: peering at the accreting polar caps of the eclipsing polar SDSS J015543.40+002807.2}.
\newblock {\em \mnras} {\bf 2006}, {\em 372},~151--162,  \href{http://arxiv.org/abs/astro-ph/0607266}{{\normalfont [arXiv:astro-ph/astro-ph/0607266]}}.
\newblock {\url{https://doi.org/10.1111/j.1365-2966.2006.10834.x}}.

\bibitem[{Barnes} et~al.(2008){Barnes}, {Cottrell}, {Albrow}, {Frost}, {Graham}, {Kershaw}, {Ritchie}, {Jones}, {Sharples}, {Bramall}, {Schmoll}, {Luke}, {Clark}, {Tyas}, {Buckley}, and {Brink}]{2008SPIE.7014E..0KB}
{Barnes}, S.I.; {Cottrell}, P.L.; {Albrow}, M.D.; {Frost}, N.; {Graham}, G.; {Kershaw}, G.; {Ritchie}, R.; {Jones}, D.; {Sharples}, R.; {Bramall}, D.;  et~al., {The optical design of the Southern African Large Telescope high resolution spectrograph: SALT HRS}.
\newblock In {\em {Ground-based and Airborne Instrumentation for Astronomy II. Edited by McLean, Ian S.; Casali, Mark M. Proceedings of the SPIE, Volume 7014, article id. 70140K, 12 pp. (2008).}};  2008; Vol. 7014, {\em {Society of Photo-Optical Instrumentation Engineers (SPIE) Conference Series}}, p. 70140K.
\newblock {\url{https://doi.org/10.1117/12.788219}}.

\bibitem[{Bramall} et~al.(2010){Bramall}, {Sharples}, {Tyas}, {Schmoll}, {Clark}, {Luke}, {Looker}, {Dipper}, {Ryan}, {Buckley}, {Brink}, and {Barnes}]{2010SPIE.7735E..4FB}
{Bramall}, D.G.; {Sharples}, R.; {Tyas}, L.; {Schmoll}, J.; {Clark}, P.; {Luke}, P.; {Looker}, N.; {Dipper}, N.A.; {Ryan}, S.; {Buckley}, D.A.H.;  et~al., {The SALT HRS spectrograph: final design, instrument capabilities, and operational modes}.
\newblock In {\em {Proceedings of the SPIE, Volume 7735, id. 77354F (2010).}};  2010; Vol. 7735, {\em {Society of Photo-Optical Instrumentation Engineers (SPIE) Conference Series}}, p. 77354F.
\newblock {\url{https://doi.org/10.1117/12.856382}}.

\bibitem[{Bramall} et~al.(2012){Bramall}, {Schmoll}, {Tyas}, {Clark}, {Younger}, {Sharples}, {Dipper}, {Ryan}, {Buckley}, and {Brink}]{2012SPIE.8446E..0AB}
{Bramall}, D.G.; {Schmoll}, J.; {Tyas}, L.M.G.; {Clark}, P.; {Younger}, E.; {Sharples}, R.M.; {Dipper}, N.A.; {Ryan}, S.G.; {Buckley}, D.A.H.; {Brink}, J., {The SALT HRS spectrograph: instrument integration and laboratory test results}.
\newblock In {\em {Ground-based and Airborne Instrumentation for Astronomy IV. Proceedings of the SPIE, Volume 8446, article id. 84460A, 9 pp. (2012).}};  2012; Vol. 8446, {\em {Society of Photo-Optical Instrumentation Engineers (SPIE) Conference Series}}, p. 84460A.
\newblock {\url{https://doi.org/10.1117/12.925935}}.

\bibitem[{Crause} et~al.(2014){Crause}, {Sharples}, {Bramall}, {Schmoll}, {Clark}, {Younger}, {Tyas}, {Ryan}, {Brink}, {Strydom}, {Buckley}, {Wilkinson}, {Crawford}, and {Depagne}]{2014SPIE.9147E..6TC}
{Crause}, L.A.; {Sharples}, R.M.; {Bramall}, D.G.; {Schmoll}, J.; {Clark}, P.; {Younger}, E.J.; {Tyas}, L.M.G.; {Ryan}, S.G.; {Brink}, J.D.; {Strydom}, O.J.;  et~al., {Performance of the Southern African Large Telescope (SALT) High Resolution Spectrograph (HRS)}.
\newblock In {\em {Proceedings of the SPIE, Volume 9147, id. 91476T 14 pp. (2014).}};  2014; Vol. 9147, {\em {Society of Photo-Optical Instrumentation Engineers (SPIE) Conference Series}}, p. 91476T.
\newblock {\url{https://doi.org/10.1117/12.2055635}}.

\bibitem[{Crawford} et~al.(2010){Crawford}, {Still}, {Schellart}, {Balona}, {Buckley}, {Dugmore}, {Gulbis}, {Kniazev}, {Kotze}, {Loaring}, {Nordsieck}, {Pickering}, {Potter}, {Romero Colmenero}, {Vaisanen}, {Williams}, and {Zietsman}]{2010SPIE.7737E..25C}
{Crawford}, S.M.; {Still}, M.; {Schellart}, P.; {Balona}, L.; {Buckley}, D.A.H.; {Dugmore}, G.; {Gulbis}, A.A.S.; {Kniazev}, A.; {Kotze}, M.; {Loaring}, N.;  et~al., {PySALT: the SALT science pipeline}.
\newblock In {\em {Proceedings of the SPIE, Volume 7737, id. 773725 (2010).}};  2010; Vol. 7737, {\em {Society of Photo-Optical Instrumentation Engineers (SPIE) Conference Series}}, p. 773725.
\newblock {\url{https://doi.org/10.1117/12.857000}}.

\bibitem[{Kniazev} et~al.(2016){Kniazev}, {Gvaramadze}, and {Berdnikov}]{2016MNRAS.459.3068K}
{Kniazev}, A.Y.; {Gvaramadze}, V.V.; {Berdnikov}, L.N.
\newblock {MN48: a new Galactic bona fide luminous blue variable revealed by Spitzer and SALT}.
\newblock {\em \mnras} {\bf 2016}, {\em 459},~3068--3077,  \href{http://arxiv.org/abs/1604.03942}{{\normalfont [arXiv:astro-ph.SR/1604.03942]}}.
\newblock {\url{https://doi.org/10.1093/mnras/stw889}}.

\bibitem[{Kniazev} et~al.(2019){Kniazev}, {Usenko}, {Kovtyukh}, and {Berdnikov}]{2019AstBu..74..208K}
{Kniazev}, A.Y.; {Usenko}, I.A.; {Kovtyukh}, V.V.; {Berdnikov}, L.N.
\newblock {The MAGIC Project. I. High-Resolution Spectroscopy on Salt Telescope and the Cepheid RsNor as a Test Object}.
\newblock {\em Astrophysical Bulletin} {\bf 2019}, {\em 74},~208--220.
\newblock {\url{https://doi.org/10.1134/S199034131902010X}}.

\bibitem[{Gaia Collaboration} et~al.(2022){Gaia Collaboration}, {Vallenari}, {Brown}, {Prusti}, {de Bruijne}, {Arenou}, {Babusiaux}, {Biermann}, {Creevey}, {Ducourant}, {Evans}, {Eyer}, {Guerra}, {Hutton}, {Jordi}, {Klioner}, {Lammers}, {Lindegren}, {Luri}, {Mignard}, {Panem}, {Pourbaix}, {Randich}, {Sartoretti}, {Soubiran}, {Tanga}, {Walton}, {Bailer-Jones}, {Bastian}, {Drimmel}, {Jansen}, {Katz}, {Lattanzi}, {van Leeuwen}, {Bakker}, {Cacciari}, {Casta{\~n}eda}, {De Angeli}, {Fabricius}, {Fouesneau}, {Fr{\'e}mat}, {Galluccio}, {Guerrier}, {Heiter}, {Masana}, {Messineo}, {Mowlavi}, {Nicolas}, {Nienartowicz}, {Pailler}, {Panuzzo}, {Riclet}, {Roux}, {Seabroke}, {Sordo{\o}rcit}, {Th{\'e}venin}, {Gracia-Abril}, {Portell}, {Teyssier}, {Altmann}, {Andrae}, {Audard}, {Bellas-Velidis}, {Benson}, {Berthier}, {Blomme}, {Burgess}, {Busonero}, {Busso}, {C{\'a}novas}, {Carry}, {Cellino}, {Cheek}, {Clementini}, {Damerdji}, {Davidson}, {de Teodoro}, {Nu{\~n}ez Campos}, {Delchambre}, {Dell'Oro}, {Esquej},
  {Fern{\'a}ndez-Hern{\'a}ndez}, {Fraile}, {Garabato}, {Garc{\'\i}a-Lario}, {Gosset}, {Haigron}, {Halbwachs}, {Hambly}, {Harrison}, {Hern{\'a}ndez}, {Hestroffer}, {Hodgkin}, {Holl}, {Jan{\ss}en}, {Jevardat de Fombelle}, {Jordan}, {Krone-Martins}, {Lanzafame}, {L{\"o}ffler}, {Marchal}, {Marrese}, {Moitinho}, {Muinonen}, {Osborne}, {Pancino}, {Pauwels}, {Recio-Blanco}, {Reyl{\'e}}, {Riello}, {Rimoldini}, {Roegiers}, {Rybizki}, {Sarro}, {Siopis}, {Smith}, {Sozzetti}, {Utrilla}, {van Leeuwen}, {Abbas}, {{\'A}brah{\'a}m}, {Abreu Aramburu}, {Aerts}, {Aguado}, {Ajaj}, {Aldea-Montero}, {Altavilla}, {{\'A}lvarez}, {Alves}, {Anders}, {Anderson}, {Anglada Varela}, {Antoja}, {Baines}, {Baker}, {Balaguer-N{\'u}{\~n}ez}, {Balbinot}, {Balog}, {Barache}, {Barbato}, {Barros}, {Barstow}, {Bartolom{\'e}}, {Bassilana}, {Bauchet}, {Becciani}, {Bellazzini}, {Berihuete}, {Bernet}, {Bertone}, {Bianchi}, {Binnenfeld}, {Blanco-Cuaresma}, {Blazere}, {Boch}, {Bombrun}, {Bossini}, {Bouquillon}, {Bragaglia}, {Bramante}, {Breedt},
  {Bressan}, {Brouillet}, {Brugaletta}, {Bucciarelli}, {Burlacu}, {Butkevich}, {Buzzi}, {Caffau}, {Cancelliere}, {Cantat-Gaudin}, {Carballo}, {Carlucci}, {Carnerero}, {Carrasco}, {Casamiquela}, {Castellani}, {Castro-Ginard}, {Chaoul}, {Charlot}, {Chemin}, {Chiaramida}, {Chiavassa}, {Chornay}, {Comoretto}, {Contursi}, {Cooper}, {Cornez}, {Cowell}, {Crifo}, {Cropper}, {Crosta}, {Crowley}, {Dafonte}, {Dapergolas}, {David}, {David}, {de Laverny}, {De Luise}, {De March}, {De Ridder}, {de Souza}, {de Torres}, {del Peloso}, {del Pozo}, {Delbo}, {Delgado}, {Delisle}, {Demouchy}, {Dharmawardena}, {Di Matteo}, {Diakite}, {Diener}, {Distefano}, {Dolding}, {Edvardsson}, {Enke}, {Fabre}, {Fabrizio}, {Faigler}, {Fedorets}, {Fernique}, {Fienga}, {Figueras}, {Fournier}, {Fouron}, {Fragkoudi}, {Gai}, {Garcia-Gutierrez}, {Garcia-Reinaldos}, {Garc{\'\i}a-Torres}, {Garofalo}, {Gavel}, {Gavras}, {Gerlach}, {Geyer}, {Giacobbe}, {Gilmore}, {Girona}, {Giuffrida}, {Gomel}, {Gomez}, {Gonz{\'a}lez-N{\'u}{\~n}ez},
  {Gonz{\'a}lez-Santamar{\'\i}a}, {Gonz{\'a}lez-Vidal}, {Granvik}, {Guillout}, {Guiraud}, {Guti{\'e}rrez-S{\'a}nchez}, {Guy}, {Hatzidimitriou}, {Hauser}, {Haywood}, {Helmer}, {Helmi}, {Sarmiento}, {Hidalgo}, {Hilger}, {H{\l}adczuk}, {Hobbs}, {Holland}, {Huckle}, {Jardine}, {Jasniewicz}, {Jean-Antoine Piccolo}, {Jim{\'e}nez-Arranz}, {Jorissen}, {Juaristi Campillo}, {Julbe}, {Karbevska}, {Kervella}, {Khanna}, {Kontizas}, {Kordopatis}, {Korn}, {K{\'o}sp{\'a}l}, {Kostrzewa-Rutkowska}, {Kruszy{\'n}ska}, {Kun}, {Laizeau}, {Lambert}, {Lanza}, {Lasne}, {Le Campion}, {Lebreton}, {Lebzelter}, {Leccia}, {Leclerc}, {Lecoeur-Taibi}, {Liao}, {Licata}, {Lindstr{\o}m}, {Lister}, {Livanou}, {Lobel}, {Lorca}, {Loup}, {Madrero Pardo}, {Magdaleno Romeo}, {Managau}, {Mann}, {Manteiga}, {Marchant}, {Marconi}, {Marcos}, {Marcos Santos}, {Mar{\'\i}n Pina}, {Marinoni}, {Marocco}, {Marshall}, {Polo}, {Mart{\'\i}n-Fleitas}, {Marton}, {Mary}, {Masip}, {Massari}, {Mastrobuono-Battisti}, {Mazeh}, {McMillan}, {Messina}, {Michalik},
  {Millar}, {Mints}, {Molina}, {Molinaro}, {Moln{\'a}r}, {Monari}, {Mongui{\'o}}, {Montegriffo}, {Montero}, {Mor}, {Mora}, {Morbidelli}, {Morel}, {Morris}, {Muraveva}, {Murphy}, {Musella}, {Nagy}, {Noval}, {Oca{\~n}a}, {Ogden}, {Ordenovic}, {Osinde}, {Pagani}, {Pagano}, {Palaversa}, {Palicio}, {Pallas-Quintela}, {Panahi}, {Payne-Wardenaar}, {Pe{\~n}alosa Esteller}, {Penttil{\"a}}, {Pichon}, {Piersimoni}, {Pineau}, {Plachy}, {Plum}, {Poggio}, {Pr{\v{s}}a}, {Pulone}, {Racero}, {Ragaini}, {Rainer}, {Raiteri}, {Rambaux}, {Ramos}, {Ramos-Lerate}, {Re Fiorentin}, {Regibo}, {Richards}, {Rios Diaz}, {Ripepi}, {Riva}, {Rix}, {Rixon}, {Robichon}, {Robin}, {Robin}, {Roelens}, {Rogues}, {Rohrbasser}, {Romero-G{\'o}mez}, {Rowell}, {Royer}, {Ruz Mieres}, {Rybicki}, {Sadowski}, {S{\'a}ez N{\'u}{\~n}ez}, {Sagrist{\`a} Sell{\'e}s}, {Sahlmann}, {Salguero}, {Samaras}, {Sanchez Gimenez}, {Sanna}, {Santove{\~n}a}, {Sarasso}, {Schultheis}, {Sciacca}, {Segol}, {Segovia}, {S{\'e}gransan}, {Semeux}, {Shahaf}, {Siddiqui}, {Siebert},
  {Siltala}, {Silvelo}, {Slezak}, {Slezak}, {Smart}, {Snaith}, {Solano}, {Solitro}, {Souami}, {Souchay}, {Spagna}, {Spina}, {Spoto}, {Steele}, {Steidelm{\"u}ller}, {Stephenson}, {S{\"u}veges}, {Surdej}, {Szabados}, {Szegedi-Elek}, {Taris}, {Taylo}, {Teixeira}, {Tolomei}, {Tonello}, {Torra}, {Torra}, {Torralba Elipe}, {Trabucchi}, {Tsounis}, {Turon}, {Ulla}, {Unger}, {Vaillant}, {van Dillen}, {van Reeven}, {Vanel}, {Vecchiato}, {Viala}, {Vicente}, {Voutsinas}, {Weiler}, {Wevers}, {Wyrzykowski}, {Yoldas}, {Yvard}, {Zhao}, {Zorec}, {Zucker}, and {Zwitter}]{gaia3}
{Gaia Collaboration}.; {Vallenari}, A.; {Brown}, A.G.A.; {Prusti}, T.; {de Bruijne}, J.H.J.; {Arenou}, F.; {Babusiaux}, C.; {Biermann}, M.; {Creevey}, O.L.; {Ducourant}, C.;  et~al.
\newblock {Gaia Data Release 3: Summary of the content and survey properties}.
\newblock {\em arXiv e-prints} {\bf 2022}, p. arXiv:2208.00211,  \href{http://arxiv.org/abs/2208.00211}{{\normalfont [arXiv:astro-ph.GA/2208.00211]}}.

\bibitem[{Tkachenko}(2015)]{gssp}
{Tkachenko}, A.
\newblock {Grid search in stellar parameters: a software for spectrum analysis of single stars and binary systems}.
\newblock {\em \aap} {\bf 2015}, {\em 581},~A129,  \href{http://arxiv.org/abs/1507.02864}{{\normalfont [arXiv:astro-ph.SR/1507.02864]}}.
\newblock {\url{https://doi.org/10.1051/0004-6361/201526513}}.

\bibitem[{Zechmeister} and {K{\"u}rster}(2009)]{gls}
{Zechmeister}, M.; {K{\"u}rster}, M.
\newblock {The generalised Lomb-Scargle periodogram. A new formalism for the floating-mean and Keplerian periodograms}.
\newblock {\em \aap} {\bf 2009}, {\em 496},~577--584,  \href{http://arxiv.org/abs/0901.2573}{{\normalfont [arXiv:astro-ph.IM/0901.2573]}}.
\newblock {\url{https://doi.org/10.1051/0004-6361:200811296}}.

\bibitem[{Kovalev} et~al.(2024){Kovalev}, {Chen}, and {Han}]{j115}
{Kovalev}, M.Y.; {Chen}, X.; {Han}, Z.
\newblock {J115307.93+353528.2{\textemdash}Spectroscopic Twin Binary, Composed of Solar Like Stars. Orbital Solution from Poorly Resolved Double-line Structure}.
\newblock {\em Research Notes of the American Astronomical Society} {\bf 2024}, {\em 8},~175.
\newblock {\url{https://doi.org/10.3847/2515-5172/ad5f2f}}.

\bibitem[{Czesla} et~al.(2019){Czesla}, {Schr{\"o}ter}, {Schneider}, {Huber}, {Pfeifer}, {Andreasen}, and {Zechmeister}]{pyasl}
{Czesla}, S.; {Schr{\"o}ter}, S.; {Schneider}, C.P.; {Huber}, K.F.; {Pfeifer}, F.; {Andreasen}, D.T.; {Zechmeister}, M.
\newblock {PyA: Python astronomy-related packages},  2019,  \href{http://arxiv.org/abs/1906.010}{{\normalfont [1906.010]}}.

\bibitem[{Huang} et~al.(2020{\natexlab{a}}){Huang}, {Vanderburg}, {P{\'a}l}, {Sha}, {Yu}, {Fong}, {Fausnaugh}, {Shporer}, {Guerrero}, {Vanderspek}, and {Ricker}]{tess1}
{Huang}, C.X.; {Vanderburg}, A.; {P{\'a}l}, A.; {Sha}, L.; {Yu}, L.; {Fong}, W.; {Fausnaugh}, M.; {Shporer}, A.; {Guerrero}, N.; {Vanderspek}, R.;  et~al.
\newblock {Photometry of 10 Million Stars from the First Two Years of TESS Full Frame Images: Part I}.
\newblock {\em Research Notes of the American Astronomical Society} {\bf 2020}, {\em 4},~204,  \href{http://arxiv.org/abs/2011.06459}{{\normalfont [arXiv:astro-ph.EP/2011.06459]}}.
\newblock {\url{https://doi.org/10.3847/2515-5172/abca2e}}.

\bibitem[{Huang} et~al.(2020{\natexlab{b}}){Huang}, {Vanderburg}, {P{\'a}l}, {Sha}, {Yu}, {Fong}, {Fausnaugh}, {Shporer}, {Guerrero}, {Vanderspek}, and {Ricker}]{tess2}
{Huang}, C.X.; {Vanderburg}, A.; {P{\'a}l}, A.; {Sha}, L.; {Yu}, L.; {Fong}, W.; {Fausnaugh}, M.; {Shporer}, A.; {Guerrero}, N.; {Vanderspek}, R.;  et~al.
\newblock {Photometry of 10 Million Stars from the First Two Years of TESS Full Frame Images: Part II}.
\newblock {\em Research Notes of the American Astronomical Society} {\bf 2020}, {\em 4},~206.
\newblock {\url{https://doi.org/10.3847/2515-5172/abca2d}}.

\bibitem[{Conroy} et~al.(2020){Conroy}, {Kochoska}, {Hey}, {Pablo}, {Hambleton}, {Jones}, {Giammarco}, {Abdul-Masih}, and {Pr{\v{s}}a}]{phoebe}
{Conroy}, K.E.; {Kochoska}, A.; {Hey}, D.; {Pablo}, H.; {Hambleton}, K.M.; {Jones}, D.; {Giammarco}, J.; {Abdul-Masih}, M.; {Pr{\v{s}}a}, A.
\newblock {Physics of Eclipsing Binaries. V. General Framework for Solving the Inverse Problem}.
\newblock {\em \apjs} {\bf 2020}, {\em 250},~34,  \href{http://arxiv.org/abs/2006.16951}{{\normalfont [arXiv:astro-ph.SR/2006.16951]}}.
\newblock {\url{https://doi.org/10.3847/1538-4365/abb4e2}}.

\bibitem[{Steinmetz} et~al.(2020){Steinmetz}, {Matijevi{\v{c}}}, {Enke}, {Zwitter}, {Guiglion}, {McMillan}, {Kordopatis}, {Valentini}, {Chiappini}, {Casagrande}, {Wojno}, {Anguiano}, {Bienaym{\'e}}, {Bijaoui}, {Binney}, {Burton}, {Cass}, {de Laverny}, {Fiegert}, {Freeman}, {Fulbright}, {Gibson}, {Gilmore}, {Grebel}, {Helmi}, {Kunder}, {Munari}, {Navarro}, {Parker}, {Ruchti}, {Recio-Blanco}, {Reid}, {Seabroke}, {Siviero}, {Siebert}, {Stupar}, {Watson}, {Williams}, {Wyse}, {Anders}, {Antoja}, {Birko}, {Bland-Hawthorn}, {Bossini}, {Garc{\'\i}a}, {Carrillo}, {Chaplin}, {Elsworth}, {Famaey}, {Gerhard}, {Jofre}, {Just}, {Mathur}, {Miglio}, {Minchev}, {Monari}, {Mosser}, {Ritter}, {Rodrigues}, {Scholz}, {Sharma}, {Sysoliatina}, and {RAVE Collaboration}]{dr6rave}
{Steinmetz}, M.; {Matijevi{\v{c}}}, G.; {Enke}, H.; {Zwitter}, T.; {Guiglion}, G.; {McMillan}, P.J.; {Kordopatis}, G.; {Valentini}, M.; {Chiappini}, C.; {Casagrande}, L.;  et~al.
\newblock {The Sixth Data Release of the Radial Velocity Experiment (RAVE). I. Survey Description, Spectra, and Radial Velocities}.
\newblock {\em \aj} {\bf 2020}, {\em 160},~82,  \href{http://arxiv.org/abs/2002.04377}{{\normalfont [arXiv:astro-ph.SR/2002.04377]}}.
\newblock {\url{https://doi.org/10.3847/1538-3881/ab9ab9}}.

\bibitem[{Birko} et~al.(2019){Birko}, {Zwitter}, {Grebel}, {Parker}, {Kordopatis}, {Bland-Hawthorn}, {Freeman}, {Guiglion}, {Gibson}, {Navarro}, {Reid}, {Seabroke}, {Steinmetz}, and {Watson}]{sb1rave2}
{Birko}, D.; {Zwitter}, T.; {Grebel}, E.K.; {Parker}, Q.A.; {Kordopatis}, G.; {Bland-Hawthorn}, J.; {Freeman}, K.; {Guiglion}, G.; {Gibson}, B.K.; {Navarro}, J.;  et~al.
\newblock {Single-lined Spectroscopic Binary Star Candidates from a Combination of the RAVE and Gaia DR2 Surveys}.
\newblock {\em \aj} {\bf 2019}, {\em 158},~155,  \href{http://arxiv.org/abs/1906.11486}{{\normalfont [arXiv:astro-ph.SR/1906.11486]}}.
\newblock {\url{https://doi.org/10.3847/1538-3881/ab3cc1}}.

\bibitem[{Matijevi{\v{c}}} et~al.(2010){Matijevi{\v{c}}}, {Zwitter}, {Munari}, {Bienaym{\'e}}, {Binney}, {Bland-Hawthorn}, {Boeche}, {Campbell}, {Freeman}, {Gibson}, {Gilmore}, {Grebel}, {Helmi}, {Navarro}, {Parker}, {Seabroke}, {Siebert}, {Siviero}, {Steinmetz}, {Watson}, {Williams}, and {Wyse}]{sb2rave}
{Matijevi{\v{c}}}, G.; {Zwitter}, T.; {Munari}, U.; {Bienaym{\'e}}, O.; {Binney}, J.; {Bland-Hawthorn}, J.; {Boeche}, C.; {Campbell}, R.; {Freeman}, K.C.; {Gibson}, B.;  et~al.
\newblock {Double-lined Spectroscopic Binary Stars in the Radial Velocity Experiment Survey}.
\newblock {\em \aj} {\bf 2010}, {\em 140},~184--195,  \href{http://arxiv.org/abs/1006.2517}{{\normalfont [arXiv:astro-ph.SR/1006.2517]}}.
\newblock {\url{https://doi.org/10.1088/0004-6256/140/1/184}}.

\bibitem[{Abdurro'uf} et~al.(2022){Abdurro'uf}, {Accetta}, {Aerts}, {Silva Aguirre}, {Ahumada}, {Ajgaonkar}, {Filiz Ak}, {Alam}, {Allende Prieto}, {Almeida}, {Anders}, {Anderson}, {Andrews}, {Anguiano}, {Aquino-Ort{\'\i}z}, {Arag{\'o}n-Salamanca}, {Argudo-Fern{\'a}ndez}, {Ata}, {Aubert}, {Avila-Reese}, {Badenes}, {Barb{\'a}}, {Barger}, {Barrera-Ballesteros}, {Beaton}, {Beers}, {Belfiore}, {Bender}, {Bernardi}, {Bershady}, {Beutler}, {Bidin}, {Bird}, {Bizyaev}, {Blanc}, {Blanton}, {Boardman}, {Bolton}, {Boquien}, {Borissova}, {Bovy}, {Brandt}, {Brown}, {Brownstein}, {Brusa}, {Buchner}, {Bundy}, {Burchett}, {Bureau}, {Burgasser}, {Cabang}, {Campbell}, {Cappellari}, {Carlberg}, {Wanderley}, {Carrera}, {Cash}, {Chen}, {Chen}, {Cherinka}, {Chiappini}, {Choi}, {Chojnowski}, {Chung}, {Clerc}, {Cohen}, {Comerford}, {Comparat}, {da Costa}, {Covey}, {Crane}, {Cruz-Gonzalez}, {Culhane}, {Cunha}, {Dai}, {Damke}, {Darling}, {Davidson}, {Davies}, {Dawson}, {De Lee}, {Diamond-Stanic}, {Cano-D{\'\i}az}, {S{\'a}nchez},
  {Donor}, {Duckworth}, {Dwelly}, {Eisenstein}, {Elsworth}, {Emsellem}, {Eracleous}, {Escoffier}, {Fan}, {Farr}, {Feng}, {Fern{\'a}ndez-Trincado}, {Feuillet}, {Filipp}, {Fillingham}, {Frinchaboy}, {Fromenteau}, {Galbany}, {Garc{\'\i}a}, {Garc{\'\i}a-Hern{\'a}ndez}, {Ge}, {Geisler}, {Gelfand}, {G{\'e}ron}, {Gibson}, {Goddy}, {Godoy-Rivera}, {Grabowski}, {Green}, {Greener}, {Grier}, {Griffith}, {Guo}, {Guy}, {Hadjara}, {Harding}, {Hasselquist}, {Hayes}, {Hearty}, {Hern{\'a}ndez}, {Hill}, {Hogg}, {Holtzman}, {Horta}, {Hsieh}, {Hsu}, {Hsu}, {Huber}, {Huertas-Company}, {Hutchinson}, {Hwang}, {Ibarra-Medel}, {Chitham}, {Ilha}, {Imig}, {Jaekle}, {Jayasinghe}, {Ji}, {Johnson}, {Jones}, {J{\"o}nsson}, {Katkov}, {Khalatyan}, {Kinemuchi}, {Kisku}, {Knapen}, {Kneib}, {Kollmeier}, {Kong}, {Kounkel}, {Kreckel}, {Krishnarao}, {Lacerna}, {Lane}, {Langgin}, {Lavender}, {Law}, {Lazarz}, {Leung}, {Leung}, {Lewis}, {Li}, {Li}, {Lian}, {Liang}, {Lin}, {Lin}, {Lin}, {Lintott}, {Long}, {Longa-Pe{\~n}a}, {L{\'o}pez-Cob{\'a}}, {Lu},
  {Lundgren}, {Luo}, {Mackereth}, {de la Macorra}, {Mahadevan}, {Majewski}, {Manchado}, {Mandeville}, {Maraston}, {Margalef-Bentabol}, {Masseron}, {Masters}, {Mathur}, {McDermid}, {Mckay}, {Merloni}, {Merrifield}, {Meszaros}, {Miglio}, {Di Mille}, {Minniti}, {Minsley}, and {Monachesi}]{apogee17}
{Abdurro'uf}.; {Accetta}, K.; {Aerts}, C.; {Silva Aguirre}, V.; {Ahumada}, R.; {Ajgaonkar}, N.; {Filiz Ak}, N.; {Alam}, S.; {Allende Prieto}, C.; {Almeida}, A.;  et~al.
\newblock {The Seventeenth Data Release of the Sloan Digital Sky Surveys: Complete Release of MaNGA, MaStar, and APOGEE-2 Data}.
\newblock {\em \apjs} {\bf 2022}, {\em 259},~35,  \href{http://arxiv.org/abs/2112.02026}{{\normalfont [arXiv:astro-ph.GA/2112.02026]}}.
\newblock {\url{https://doi.org/10.3847/1538-4365/ac4414}}.

\bibitem[{Hwang} et~al.(2022){Hwang}, {Ting}, and {Zakamska}]{widebinary}
{Hwang}, H.C.; {Ting}, Y.S.; {Zakamska}, N.L.
\newblock {The eccentricity distribution of wide binaries and their individual measurements}.
\newblock {\em \mnras} {\bf 2022}, {\em 512},~3383--3399,  \href{http://arxiv.org/abs/2111.01789}{{\normalfont [arXiv:astro-ph.SR/2111.01789]}}.
\newblock {\url{https://doi.org/10.1093/mnras/stac675}}.

\bibitem[{Iliji{\'c}}(2017)]{fd3}
{Iliji{\'c}}, S.
\newblock {fd3: Spectral disentangling of double-lined spectroscopic binary stars}.
\newblock Astrophysics Source Code Library, record ascl:1705.012,  2017,  \href{http://arxiv.org/abs/1705.012}{{\normalfont [1705.012]}}.

\bibitem[{Kovalev} et~al.(2024){Kovalev}, {Chen}, and {Han}]{fakesb2}
{Kovalev}, M.; {Chen}, X.; {Han}, Z.
\newblock {Spectroscopic triples and a chance alignment: a solution for the problem of suspicious mass ratios for SB2s from Wilson method}.
\newblock {\em \mnras} {\bf 2024}, {\em 527},~346--355,  \href{http://arxiv.org/abs/2310.09030}{{\normalfont [arXiv:astro-ph.SR/2310.09030]}}.
\newblock {\url{https://doi.org/10.1093/mnras/stad3185}}.

\bibitem[{Ilijic} et~al.(2004){Ilijic}, {Hensberge}, {Pavlovski}, and {Freyhammer}]{fd3norm}
{Ilijic}, S.; {Hensberge}, H.; {Pavlovski}, K.; {Freyhammer}, L.M.
\newblock {Obtaining normalised component spectra with FDBinary}.
\newblock In Proceedings of the Spectroscopically and Spatially Resolving the Components of the Close Binary Stars; {Hilditch}, R.W.; {Hensberge}, H.; {Pavlovski}, K., Eds.,  2004, Vol. 318, {\em Astronomical Society of the Pacific Conference Series}, pp. 111--113.

\bibitem[{Kovalev} et~al.(2025){Kovalev}, {Matekov}, {Guo}, {Chen}, and {Han}]{ttau}
{Kovalev}, M.; {Matekov}, A.; {Guo}, S.; {Chen}, X.; {Han}, Z.
\newblock {The new compact triple system: discovery of bright third star around contact binary using LAMOST-MRS spectra and photometry}.
\newblock {\em \mnras} {\bf 2025}, {\em 539},~3830--3841,  \href{http://arxiv.org/abs/2504.21287}{{\normalfont [arXiv:astro-ph.SR/2504.21287]}}.
\newblock {\url{https://doi.org/10.1093/mnras/staf738}}.

\bibitem[{Balona}(2019)]{balona}
{Balona}, L.A.
\newblock {Evidence for spots on hot stars suggests major revision of stellar physics}.
\newblock {\em \mnras} {\bf 2019}, {\em 490},~2112--2116,  \href{http://arxiv.org/abs/1910.01584}{{\normalfont [arXiv:astro-ph.SR/1910.01584]}}.
\newblock {\url{https://doi.org/10.1093/mnras/stz2808}}.

\bibitem[{Andersen} and {Nordstrom}(1983)]{1983A&AS...52..471A}
{Andersen}, J.; {Nordstrom}, B.
\newblock {Radial velocities of bright southern stars. I. 139 B-type HR and FK stars.}
\newblock {\em \aaps} {\bf 1983}, {\em 52},~471--478.

\bibitem[{Wenger} et~al.(2000){Wenger}, {Ochsenbein}, {Egret}, {Dubois}, {Bonnarel}, {Borde}, {Genova}, {Jasniewicz}, {Lalo{\"e}}, {Lesteven}, and {Monier}]{simbad}
{Wenger}, M.; {Ochsenbein}, F.; {Egret}, D.; {Dubois}, P.; {Bonnarel}, F.; {Borde}, S.; {Genova}, F.; {Jasniewicz}, G.; {Lalo{\"e}}, S.; {Lesteven}, S.;  et~al.
\newblock {The SIMBAD astronomical database. The CDS reference database for astronomical objects}.
\newblock {\em \aaps} {\bf 2000}, {\em 143},~9--22,  \href{http://arxiv.org/abs/astro-ph/0002110}{{\normalfont [arXiv:astro-ph/astro-ph/0002110]}}.
\newblock {\url{https://doi.org/10.1051/aas:2000332}}.

\end{thebibliography}
\bibliographystyle{Definitions/mdpi.bst}

%


\PublishersNote{}
\end{adjustwidth}
\end{document}